\documentclass[aps,prx,showpacs,twocolumn,reprint,superscriptaddress,nofootinbib]{revtex4-2}

\usepackage[dvips]{graphicx}
\usepackage{amsmath,amssymb,amsthm,mathrsfs,amsfonts,dsfont,mathtools,physics}
\usepackage{epsfig}
\usepackage{braket}

\usepackage[dvipsnames]{xcolor}

\usepackage{placeins}
\usepackage{bm}
\usepackage{enumerate}
\usepackage{color}
\usepackage{graphicx}
\usepackage{textcomp}
\usepackage{float}
\usepackage[version=4]{mhchem}
\usepackage{yquant, quantikz}
\useyquantlanguage{groups}
\usetikzlibrary{calc}
\usepackage{pgfplots}
\pgfplotsset{compat=newest}

\usepackage{enumitem}

\usepackage[normalem]{ulem}
\usepackage{comment}
\usepackage{xcolor}
\usepackage{silence}
\usepackage{amsthm}

\PassOptionsToPackage{breaklinks}{hyperref}
\usepackage{xurl}
\usepackage{hyperref}
\hypersetup{
    colorlinks,
    linkcolor={blue!50!black},
    citecolor={blue!50!black},
    urlcolor={blue!80!black}
}
\usepackage[capitalize]{cleveref}

\Crefname{theorem}{Theorem}{Theorems}
\theoremstyle{remark}

\newcommand{\qmaddress}{\affiliation{Quantum Motion, 9 Sterling Way, London N7 9HJ, United Kingdom}}
\newcommand{\oxaddress}{\affiliation{Department of Materials, University of Oxford, Parks Road, Oxford OX1 3PH, United Kingdom}}
\newcommand{\GSaddress}{\affiliation{Goldman Sachs, New York, NY, USA}}
\newcommand{\Bristoladdress}{\affiliation{School of Mathematics, University of Bristol, Fry Building, Woodland Road, 
Bristol,
BS8 1UG, United Kingdom}}

\makeatletter

\DeclareRobustCommand\rvdots{%
\vbox{%
\baselineskip4\p@\lineskiplimit\z@%
\kern-\p@%
\hbox{.}\hbox{.}\hbox{.}%
}%
}

\begin{document}


\title{Low Depth Phase Oracle Using a Parallel Piecewise Circuit}
\author{Zhu Sun}
\email[zhu.sun@exeter.ox.ac.uk]{}
\qmaddress
\oxaddress

\author{Gregory Boyd}
\qmaddress
\author{Zhenyu Cai}
\qmaddress
\author{Hamza Jnane}
\qmaddress
\oxaddress
\author{B\'alint Koczor}
\qmaddress
\affiliation{Mathematical Institute, University of Oxford, Woodstock Road, Oxford OX2 6GG, United Kingdom}
\author{Richard Meister}
\qmaddress
\affiliation{Department of Computing, Imperial College London, South Kensington Campus, London SW7 2AZ, United Kingdom}
\author{Romy Minko}
\qmaddress
\Bristoladdress
\author{Benjamin Pring}
\qmaddress
\Bristoladdress
\author{Simon C. Benjamin}
\qmaddress
\oxaddress
\author{Nikitas Stamatopoulos}
\GSaddress

\begin{abstract}
    We explore the important task of applying a phase $\exp(i\,f(x))$ to a computational basis state $\ket{x}$. The closely related task of rotating a target qubit by an angle depending on $f(x)$ is also studied. Such operations are key in many quantum subroutines, and frequently $f(x)$ can be well-approximated by a piecewise function; examples range from the application of diagonal Hamiltonian terms (such as the Coulomb interaction) in grid-based many-body simulation, to derivative pricing algorithms. Here we exploit a parallelisation of the piecewise approach so that all constituent elementary rotations are performed simultaneously, that is, we achieve a total rotation depth of one. Moreover, we explore the use of recursive catalyst `towers' to implement these elementary rotations efficiently. 
    We find that strategies prioritising execution speed can
    achieve circuit depth as low as $O(\log{n}{+}\log{S})$ for a register of $n$ qubits and a piecewise approximation of $S$ sections (presuming prior preparation of enabling resource states), albeit total qubit count then scales with $S$. In the limit of multiple repetitions of the oracle, we find that catalyst tower approaches have an $O(S\cdot n)$ T-count.
    
\end{abstract}

\maketitle

\section{Introduction}
The speedup promised by quantum computing has drawn intense attention over the last few decades. Quantum algorithms are explored for problems in various research fields including cryptography \cite{shor1999polynomial}, linear systems of equations \cite{PhysRevLett.103.150502}, optimisation \cite{hogg2000quantum} and computational chemistry \cite{RevModPhys.92.015003}, among others. In these studies, algorithms usually start by assuming the ability, often granted by an `oracle', to encode some function value $f(x)$ related to $\ket{x}$ into a quantum state. Efficient implementation of such oracles is extensively studied in the context of e.g. state preparation. 

In this work, we study the common task of encoding a function value into phase, i.e. 
\begin{equation}
\ket{x}\mapsto\exp(i f(x))\ket{x},    
\end{equation}
and we propose an efficient circuit implementation for this type of oracle. With the addition of a few Clifford gates, one instead obtains a variant which implements 
\begin{equation}
\ket{x}\ket{0}\mapsto\ket{x}\left(\,\cos{f(x)}\ket{0}+i \sin{f(x)}\ket{1}\,\right). 
\end{equation}

For either case, we use classical pre-computation to make a piecewise approximation of the function $f(x)$, avoiding expensive arithmetic. By combining fan-out techniques with a repeat-until-success scheme, the circuit depth can be as low as $O(\log_2 {n})$ for an $n$-qubit input register. We also employ `catalyst towers' to moderate the high demand on magic states, which is a common problem for conventional methods. 

Our approach is beneficial for well-behaved functional forms and is particularly beneficial when the circuit is used as a repeatedly applied subroutine, such as time evolution, amplitude amplification, etc. A good example is the $1/r$ Coulomb potential. It has a simple functional form while being crucial to quantum chemistry simulations, which is one of the three application examples we present. Conversely, there are of course certain functions that are impractical for piecewise approximation (see \cref{appx:heuristic}), for example certain oscillatory functions used to benchmark optimization algorithms, e.g. the Ackley function \cite{ackley2012connectionist}.

Our focus is on minimising depth while maintaining an efficient use of resources such as magic states; to achieve this, we allow ourselves to increase the number of qubits that are concurrently active; essentially, a space-time trade-off. Consequently, the methods may be particularly relevant to quantum computing platforms where there is the hope of large numbers of qubits, such as silicon spin chip-based devices~\cite{siPlatform}. For optimisation against other priorities, we point readers toward the complementary discussion in \cite{huangApproximateRealtimeEvolution2024}, which compares methods of applying a phase focusing on minimal-width implementations. 

Historically, the word `oracle' has been widely used in quantum algorithms studies (as early as e.g. Shor's algorithm \cite{shor1999polynomial} and Grover's algorithm \cite{grover1996fast}) to describe the black box which gives access to some function $f(x)$. Thus, in the context of algorithmic complexity, its construction details are irrelevant. Since this work discusses the exact implementations, we narrow down the definition of `oracle': a circuit that encodes some function $f(x)$ into a quantum state, and $f(x)$ must be known so that it can be approximated in a piecewise fashion. This then excludes the functions in e.g. Shor's algorithm and Grover's algorithm, simply because knowing the relevant functions would directly solve the problem. 

The rest of the paper is organised as follows: in \cref{sec:prior} we discuss the prior work. The main implementation is explained in \cref{sec:parallelising}, followed by the options for the key part of our implementation in \cref{sec:options}, in which we introduce the catalyst towers. The cost analysis for different options can be found in \cref{sec:cost}. In \cref{sec:examples}, we provide three example applications which can benefit from our work, namely financial derivatives pricing, quantum chemistry and quantum simulation.

\section{Prior Work} \label{sec:prior}
The need of encoding a function onto a quantum state goes back to the very early days of quantum computing and it is essential in many quantum computing applications. Depending on the problem, one might want to encode the information into 

\begin{itemize}
    \item the register, $\ket{x}\ket{0}\mapsto\ket{x}\ket{f(x)}$, e.g. Simon's problem \cite{simon1997power},
    \item the amplitude, \newline$\ket{x}\!\ket{0}\!\mapsto\!\ket{x}\!(\sqrt{f(x)}\!\ket{\psi}{+}\sqrt{1{-}f(x)}\!\ket{\psi^{\perp}})$ (for some normalised $\ket{\psi}$ and orthogonal $\ket{\psi^{\perp}}$), e.g. quantum optimisation \cite{gilyen2019optimizing}, or
    \item the phase, $\ket{x}\mapsto\exp(i f(x))\ket{x}$, e.g. quantum phase estimation \cite{kitaev1995quantum}.
\end{itemize}

\noindent \textbf{Oracle models.} For solving some specific problems in an application area, the three aforementioned models are usually cast into oracles and the related algorithms are evaluated by the query complexity. These three models of oracles are often closely related to each other. For example, in \cite{gilyen2019optimizing} the authors show that given an amplitude oracle, it can be used to simulate the corresponding phase oracle to $\epsilon$ precision with $O(\log(1/\epsilon))$ queries and the conversion from phase oracle back to amplitude oracle also requires logarithmic overheads in the precision.\\

\noindent \textbf{State preparation methods.} The encoding of a function into amplitudes is generalised in the context of quantum state preparation, which seeks to prepare a state of the form $\sum_x f(x)/N_f\ket{x}$, where $N_f$ is a normalising constant. As a standard technique, the key step in the Grover-Rudolph method \cite{grover2002creating} first writes the bit string into an ancillary register using  arithmetic. Applying rotations to the ancillae controlled on the bit strings will then transform the function value into the amplitude. Later methods like quantum rejection sampling \cite{ozols2013quantum} and quantum eigenvalue transformation \cite{mcardle2022quantum} seek to address the same problem without the expensive arithmetic. Moreover, if the controlled-rotations in the Grover-Rudolph method are replaced by a phase $Ph(\theta)=\text{diag}(1,\exp(i \theta))$ on each bit in the string, then it implements a phase oracle. Again, the use of arithmetic is significant  in the total cost of the method.\\

\noindent \textbf{The QROM approaches.} The existence of the oracle $U_f\ket{x}\ket{0}\mapsto\ket{x}\ket{f(x)}$ has been assumed since the toy-problem era of quantum algorithm. One circuit implementation of it is the Quantum Read-Only Memory (QROM) \cite{PhysRevX.8.041015} and it is widely used in quantum chemistry for e.g. state preparation \cite{Berry2019qubitizationof,PRXQuantum.2.030305}. However, the use of this type of QROM is sometimes limited by its rapid scaling in T-count and T-depth \cite{berry2023quantum}. Note that a more general model Quantum Random Access Memory (QRAM) \cite{PhysRevLett.100.160501} is frequently used to do the same task, except in its most rigorous definition, QRAM must also allow data to be written to it. A detailed explanation of the subspecies of QRAM can be found in \cite{jaques2023qram}. 

QROM approaches have also been used for the phase oracle \cite{sandersCompilationFaultTolerantQuantum2020}, first by loading information about the linear interpolation of the function, $\tilde f(x)$, and then transforming that information into a phase. This is done by calculating flags to obtain the region that contains the value of $x$, and for that region output the slope and interpolation for the linear approximation. The calculation of $\ket{\tilde f(x)}$ can then be performed with a single multiplication and addition. This register can then be used to apply a phase either by applying a rotation blocks similar to those discussed in \cref{sec:parallelising} or by adding this register into a phase-gradient state in order to apply $e^{i\tilde f(x)}$ via phase kickback \cite{Gidney2018halvingcostof}. The multiplication of $x$ with the slope will likely have the largest cost in this approach, with methods requiring either logarithmic-depth with an $O(n^{\log_2(3)})$ overhead \cite{jangQuantumBinaryField2023}, or $O(\log^2 n)$ depth with a logarithmic space overhead \cite{nieQuantumCircuitDesign2023}. For the purposes of comparing the most optimal-depth methods in this work, we will primarily compare to the Karatsuba multiplication with $O(n^{\log_2(3)})$ $T$-cost \cite{jangQuantumBinaryField2023}, although we note that when the additional cost of this method becomes relevant, one could always switch to the alternative scheme with $O(n\log n)$ cost and $\log^2 n$ depth \cite{nieQuantumCircuitDesign2023}. Once this multiplication has taken place, the subsequent addition of the intercept and application of the phase must be performed before uncomputing the arithmetic, resulting in a larger depth than some of the alternatives presented in this work. However, due to the efficiency in T-count of reusing a phase gradient state to apply a phase (only a single addition that does not depend on $\epsilon$), this method sees savings in this respect for algorithms requiring a large number of iterations of this oracle, we provide some further comments and comparison to the independent towers method in \cref{sec:options} and \cref{appx:qrom_comments}.\\

\noindent \textbf{Quantum dynamics approaches.} An eigenstate $\ket{\phi}$ of a Hamiltonian $H$ under time evolution goes through the transformation $\ket{\phi}\mapsto\exp(-iEt)\ket{\phi}$ where $E$ is the eigenvalue. The Hamiltonian simulation addresses the problem that given some $H$, its action on some state $\ket{\psi}=\sum_j c_j\ket{\phi_j}$ under time evolution is simulated by a unitary $U$ such that $U\ket{\psi} \sim \sum_j c_j\exp(-i E_j t)\ket{\phi_j}$. Relevant methods such as Trotterization \cite{Suzuki1976GeneralizedTF}, quantum signal processing \cite{low2019hamiltonian} and quantum walk \cite{berry2009black} are extensively explored.\\

\begin{figure*}
{\centering
\begin{tikzpicture}
\begin{yquant}[plusctrl/.style={/yquant/every
control/.style={/yquant/operators/every not}, /yquant/every
positive control/.style={}}]

qubit {$\ket{x_{n-1}}$} a1;
qubit {$\ket{x_{n-2}}$} a2;
nobit a3;
qubit {$\ket{x_{0}}$}  a4;
qubit {$\ket{0}$} a5;
[blue, control style=blue]
h a5;

nobit b0[2];
qubit {$\ket{0}$} b1[2];
nobit b2;
qubit {$\ket{0}$} b3;
qubit {$\ket{0}$} b4;

nobit c0[2];
qubit {$\ket{0}$} c1[2];
nobit c2;
qubit {$\ket{0}$} c3;
qubit {$\ket{0}$} c4;

nobit d0[6];
qubit {$\ket{0}$} d1[2];
nobit d2;
qubit {$\ket{0}$} d3;
qubit {$\ket{0}$} d4;

text {$\rvdots$} a3;
text {$\rvdots$} b2;
text {$\rvdots$} c2;
text {$\rvdots$} d0[2,3];
text {$\rvdots$} d2;

cnot b1[0], c1[0], d1[0] | a1;
cnot b1[1], c1[1], d1[1] | a2;
cnot b3, c3, d3 | a4;
[blue, control style=blue]
cnot b4, c4, d4 | a5;

hspace {12mm} a3;
text {$\rvdots$} a3;
text {$\qquad \; \ $} a3;
[blue, control style=blue]
cnot a1,a2,a4 | a5;
box {$\Xi_z \!(\overset{S}{\underset{i=1} \Sigma} \! \frac{\alpha_i}{2}\!)$} (a1, a2, a3, a4);
box {${\scriptstyle\! R_z(-\!\overset{S}{\underset{i=1} \Sigma}\!\gamma_i)}$} a5;
[blue, control style=blue]
cnot a1,a2,a4 | a5;
text {$\qquad \ \;$} a3;
text {$\rvdots$} a3;

text {$\rvdots$} b2;
[plusctrl] box {$flag\,1$} (b1[0],b1[1],b2,b3)|b4;
cnot b1[0,1],b3 | b4;
box {$\Xi_z (-\frac{\alpha_1}{2})$} (b1[0],b1[1],b2,b3);
hspace {1.5mm} b4;
box {$R_z(\gamma_1)$} b4;
cnot b1[0,1],b3 | b4;
[plusctrl] box {$flag\,1$} (b1[0],b1[1],b2,b3)|b4;
text {$\rvdots$} b2;

text {$\rvdots$} c2;
[plusctrl] box {$flag\,2$} (c1[0],c1[1],c2,c3)|c4;
cnot c1[0,1],c3 | c4;
box {$\Xi_z (-\frac{\alpha_2}{2})$} (c1[0],c1[1],c2,c3);
hspace {1.5mm} c4;
box {$R_z(\gamma_2)$} c4;
cnot c1[0,1],c3 | c4;
[plusctrl] box {$flag\,2$} (c1[0],c1[1],c2,c3)|c4;
text {$\rvdots$} c2;

text {$\rvdots$} d0[2,3];
hspace {23mm} d0[2,3];
text {$\rvdots$} d0[2,3];
hspace {22mm} d0[2,3];
text {$\rvdots$} d0[2,3];
text {$\rvdots$} d2;
[plusctrl] box {$flag\,S$} (d1[0],d1[1],d2,d3)|d4;
cnot d1[0,1],d3 | d4;
box {$\Xi_z (-\frac{\alpha_S}{2})$} (d1[0],d1[1],d2,d3);
hspace {1.5mm} d4;
box {$R_z(\gamma_S)$} d4;
cnot d1[0,1],d3 | d4;
[plusctrl] box {$flag\,S$} (d1[0],d1[1],d2,d3)|d4;
text {$\rvdots$} d2;

[blue, control style=blue]
cnot b4, c4, d4 | a5;
cnot b3, c3, d3 | a4;
cnot b1[1], c1[1], d1[1] | a2;
cnot b1[0], c1[0], d1[0] | a1;
[blue, control style=blue]
h a5;
text {$\rvdots$} a3;
text {$\rvdots$} b2;
text {$\rvdots$} c2;
text {$\rvdots$} d0[2,3];
text {$\qquad \qquad \rvdots$} d2;

output {$= \begin{cases}
                { e^{i f(x)}\ket{0} \ket{x}}  \\
                \color{blue}{(\cos{f(x)}\ket{0}{+}i \sin{f(x)}\ket{1})\ket{x}}
                \end{cases}$} (a1,a2,a3,a4,a5);
output {$\ket{0}$} b1[0,1],b3,b4;
output {$\ket{0}$} c1[0,1],c3,c4;
output {$\ket{0}$} d1[0,1],d3,d4;

\end{yquant}
\end{tikzpicture}
\par}
\caption{A circuit parallelising all elementary rotations. The central rotation blocks are defined by $\Xi_z (a) := \bigotimes_{i=0}^{n-1} R_z(2^i a)$, i.e. a set of single-qubit rotations performed in parallel. 
If the blue gates are implemented, then the top group of $n+1$ qubits is left in the state $(\cos{f(x)}\ket{0}+i \sin{f(x)}\ket{1})\ket{x}$ (which is also in blue in the output); otherwise the group is left in state $e^{i f(x)}\ket{0} \ket{x}$. Flag blocks flip their target qubit iff the controlling input is in the corresponding interal of the piecewise function, as explained in the text. Further optimisations to this circuit are possible (e.g. the fan-in/out), we provide details in \cref{appx:logicaland}. Both black and blue variants were verified by exact simulation with QuESTlink; this code is available from the authors on request.}
\label{fig:mainCirc}
\end{figure*}
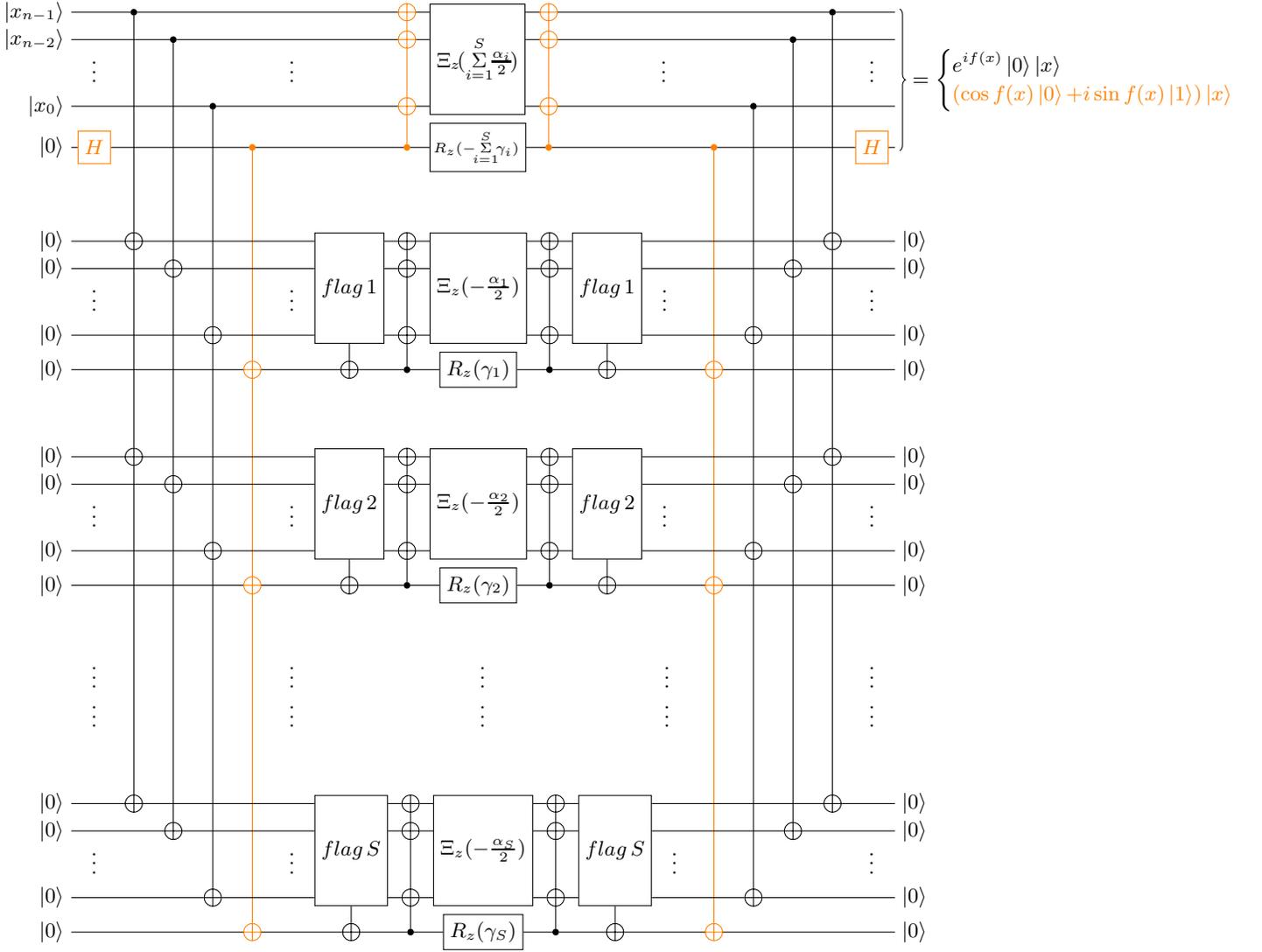

\noindent \textbf{Continuous angle rotation.}
For state preparation, continuous angle rotations are ubiquitous and this is usually expensive. Given a universal basis set of unitary gates for implementation, such as the Clifford+T gate set, direct gate synthesis leverages the underlying number theoretic properties of the gate set to determine, classically and deterministically, a sequence of basis gates that will approximate the target rotation to within the desired accuracy $\epsilon.$ For single-qubit diagonal rotation gates, as in \cref{fig:mainCirc}, this can be done optimally without using ancilla qubits for a T-depth and T-count of $O(3\log_2(1/\epsilon))$, using Clifford+T gates \cite{RossSelinger2014}. Additional techniques, such as fallback protocols \cite{BocharovRoettelerSvore2014}, probabilistic mixing \cite{Campbell2017,Hastings2017} and a combination of the two \cite{kliuchnikov2023shorter}, achieve further reductions in T depth via using ancilla qubits and projective measurements to leading order of $\log_2(1/\epsilon)$, $1.5\log_2(1/\epsilon)$ and $0.5\log_2(1/\epsilon)$, respectively. Gate synthesis protocols using smaller angle rotations instead of T gate, e.g. employing higher orders in the Clifford hierarchy, were also explored in the literature \cite{PhysRevA.91.042315,campbell2016efficient}. However, the dependence of cost on rotation accuracy can be problematic for practical purposes. We detail two alternative approaches for implementing a group of rotations using an arrangement of generalised rotation catalyst circuits~\cite{gidney2019efficient}, which we call \emph{catalyst towers}.

\section{Parallelising the piecewise function} \label{sec:parallelising}
Piecewise polynomial approximation is commonly used in mathematics and it is also not new to quantum algorithms \cite{vazquez2022enhancing,haner2018optimizing, sandersCompilationFaultTolerantQuantum2020}. However, one of its main problems is the great circuit depth due to the serial rotations. Here the solution we study is to parallelise the circuit using fan-out techniques, which is a width-depth trade-off. This is valuable for algorithms that require time optimisation, e.g. the derivative pricing algorithm example in \cref{sec:examples}. 

To parallelise, we propose the circuit in \cref{fig:mainCirc}. First, let us disregard the gates in blue to apply the phase $e^{if(x)}$ to the top register of $n$ qubits (we will come back to these blue gates later). Suppose the target function $f(x)$ is sectioned into $S$ pieces by some algorithm (we discuss one method in \cref{appx:heuristic}), and each piece $i$ is approximated by a linear function $\alpha_i x+\beta_i$, for $x$ belonging to the $i$th interval. For simplicity, define $\gamma_i=\alpha_i \phi+\beta_i$, with $\phi=(2^n-1)/2$. Note that although we use linear segments here for illustration, higher order polynomial approximations can also be used. The corresponding circuit has $S+1$ registers of $n+1$ qubits. (Strictly speaking, only $n$ qubits are used in the uppermost part of Fig.~\ref{fig:mainCirc} when the blue gates are not employed; however we include this redundant qubit to keep resource expressions compact and applicable to both black an blue variants). The input state $\ket{x} = \ket{x_0}...\ket{x_{n-1}}$ is first copied to the other S registers to parallelise the rotations. The circuit then computes the flags $1,2,...,S$ to see if the value $x$ is in section $1,2,...,S$, respectively. The exact depth of a flag block depends on the choice of section boundaries; in the examples we consider here, the flag block is equivalent to a single $\lceil \log_2{S} \rceil$-controlled Toffoli gate (the method in \cref{appx:heuristic} uses a binary sectioning). The flags are followed by the large rotation blocks, denoted as $\Xi_z (a) := \bigotimes_{i=0}^{n-1} R_z(2^i a)$ or diagrammatically as

\begin{equation}
\label{fig:rot block}
\begin{yquantgroup}
    \registers{
        qubit {} a[2];
        nobit b;
        qubit {} c;}
    \circuit{
        text {$\rvdots$} b;
        box {$\Xi_z (a)$} (a,b,c);
        text {$\rvdots$} b;}
    \equals
    \circuit{
        text {$\quad \quad \rvdots$} b;
        hspace {3.8mm} a[0];
        box {$R_z(a)$} a[0];
        hspace {3mm} a[1];
        box {$R_z(2a)$} a[1];
        box {$R_z(2^{n-1}a)$} c;}
\end{yquantgroup}
\end{equation}
where $R_z(a)=$ diag$(e^{-ia/2},e^{ia/2})$.

\begin{figure*}
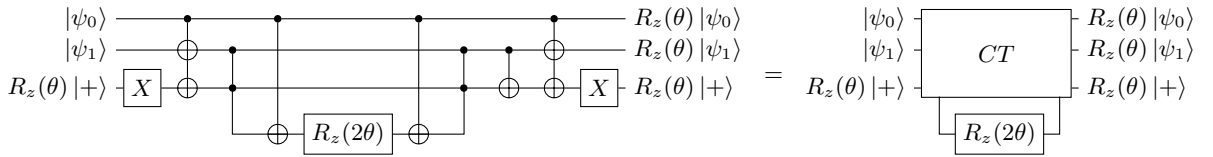

    \centering
\begin{yquantgroup}
    \registers{
        qubit {} a;
        qubit {} b;
        qubit {} c;
        qubit {} d;}
    \circuit{
        init {$\ket{\psi_0}$} a;
        init {$\ket{\psi_1}$} b;
        init {$R_z(\theta)\ket{+}$} c;
        discard d;

        x c;
        cnot b,c|a;
        zz (b,c);
        cnot d|a;
        [name=ct]
        box {$R_z(2\theta)$} d;
        cnot d|a;
        zz (b,c);
        cnot c|b;
        cnot b,c|a;
        x c;
        
        output {$R_z(\theta)\ket{\psi_0}$} a;
        output {$R_z(\theta)\ket{\psi_1}$} b;
        output {$R_z(\theta)\ket{+}$} c;
        \draw (ct)+(-1.54,0.65) |-  (ct) ;
        \draw (ct) -| + (1.54,0.65) ;}
        \equals
        \circuit{
        init {$\ket{\psi_0}$} a;
        init {$\ket{\psi_1}$} b;
        init {$R_z(\theta)\ket{+}$} c;
        
        discard d;
        hspace {4.5mm} d;
        box {$\qquad CT \qquad$} (a,b,c);
        [name=ct]
        box {$R_z(2\theta)$} d;

        output {$R_z(\theta)\ket{\psi_0}$} a;
        output {$R_z(\theta)\ket{\psi_1}$} b;
        output {$R_z(\theta)\ket{+}$} c;
        
        \draw (ct)+(-0.8,0.5) |-  (ct) ;
        \draw (ct) -| + (0.8,0.5) ;}
\end{yquantgroup}
\caption{Circuit for a 1-layer tower. The `corners'' notation for AND gate follows \cite{Gidney2018halvingcostof}, reproduced in \cref{fig:AND}.}
\label{fig:1layer}
\end{figure*}

If the flag computation for the $j$th register indicates that the value of $x$ is in section $j$, then the final qubit of that register, initially in state $\ket{0}$, is flipped to $\ket{1}$. The CNOT gates immediately preceding the rotation block are therefore activated, which effectively changes the sign on the phases applied by the rotation block. Observe, firstly, that the desired angles of $\alpha_j$ are halved and, secondly, the phases associated with the other $S-1$ sections are still applied. This is dealt with by the rotation block acting on the first register. Here, the rotation angle is the sum of each of the angles in the $S$ ancilla registers. Note that this block does not have a minus sign. Thus, the effect of this block is to cancel out the rotations applied on all registers for which the corresponding flag is not activated, and applies the remaining half of the rotations for the activated one. The single qubit $\gamma_i$ rotation offsets on the last qubit of each register are dealt with in a similar manner, via the single qubit rotations of $-\sum_{i=1}^S\gamma_i$ on the final qubit of the top register. After further disentangling the registers, the first $n$ qubits are in state $e^{i f(x)}\ket{x}$. We also provide a variant of this circuit: if the blue gates are implemented, then the information encoded in phase is transformed into amplitude, such that the top group of $n+1$ qubits end up in the state $(\cos{f(x)}\ket{0}+i \sin{f(x)}\ket{1})\ket{x}$.

The parallelisation greatly reduces the circuit depth, but there remains another problem: the huge demand for magic states required to synthesise a large number of rotations in the `large rotation block'' in \cref{fig:rot block}. One contribution of this work is exploring various constructions for applying these rotations efficiently under different usage scenarios. In particular, we focus on the improvement in T gate cost when the circuit is repeatedly applied, e.g. as a subroutine in amplitude estimation. We will describe several options for this in the next section.

\section{Options for implementing phase application} \label{sec:options}
The options we explore for implementing the phase application are:
\begin{enumerate}
\item Canonical gate synthesis (and a variant) using e.g. Clifford+T,
\item Deterministic in-circuit catalyst towers, and
\item Independent catalyst towers generating resource states, which are consumed via gate teleportation.
\item Loading a linear interpolation via QROM and using phase-kickback to apply the rotation
\end{enumerate}

Two of these methods involve `catalyst towers' whose purpose is to  generate multiple rotations, or rotation-resources, from a single input rotation and a frugal set of T-gates. The towers leverage the `generalized phase catalysis circuit' from \cite{gidney2019efficient}, and we reproduce that circuit in \cref{fig:1layer} (left). In the present context we regard this as a 1-layer tower. The $R_Z(\theta)\ket{+}$ state in the circuit is a so-called catalyst state. This acts analogously to a chemical catalyst in a reaction, in the sense that it makes the circuit less costly and remains intact after the process. For the first run of a tower circuit, the catalyst state must be synthesised in some standard way. Thus the use of such circuits in our towers is useful in the scenario that the oracle much be applied many times, or over many `rounds', to take full advantage of the reusable nature of the catalyst.

There would be no advantage in the event that the oracle were used only once; since the logical AND costs 4 T states, the first round is always more expensive than canonical gate synthesis. However, from the second round onward one only needs to synthesise the seed rotation $R_Z(2\theta)$, plus 4 extra T states for the logical AND to apply two $R_Z(\theta)$ rotations. The higher cost of the first round is averaged over the repetitions, and we will see that the break-even point for T-count is quickly reached. For simplicity, we denote this circuit as a `CT block' as shown in \cref{fig:1layer} (right).

The key to the in-circuit catalyst tower approach is the composition of the 1-layer catalyst towers. We show an example of a 2-layer in-circuit tower in \cref{fig:2layer}, and higher towers can be generalised based on this. To implement the rotations in \cref{fig:rot block}, replace the large rotation block with an $n$-layer tower, and then the desired rotation gates are applied directly on the target qubits (with one unused rotation). However, note that the scaling of the circuit depth of an $n$-layer tower is linear in $n$, this can be seen from the `V-shape' of the gates of the corresponding circuit diagram of the 2-layer tower (\cref{fig:2layer}), as shown in \cref{fig:Vshape}. Thus, this would have a greater T-depth than the direct-synthesis method --- albeit with a reduced T-cost. 

We note that a similar `in-circuit tower' is also discussed in \cite{wang2023option}. In that paper, the authors are interested in an option pricing task and optimise the conventional circuit of block-encoding of the sine function, where the parallel rotations $\bigotimes_{i=0}^{n-1} R_z(2^i a)$ are also required.

\begin{figure}[h]
    \centering
\begin{tikzpicture}
\begin{yquant}
qubit {$\ket{\psi_0}$} a;
qubit {$\ket{\psi_1}$} b;
qubit {$R_z(2^k\theta)\ket{+}$} c;
nobit d;
qubit {$\ket{\psi_2}$} e; 
qubit {$R_z(2^{k+1}\theta)\ket{+}$} f;
nobit g;

box {$\qquad\qquad CT \qquad\qquad$} (a,b,c);
hspace {3.5mm} e;
[name=ct1]
box {$\qquad\quad CT \qquad\quad$} (d,e,f);
hspace {8.5mm} g;
[name=ct2]
box {$R_z(2^{k+2}\theta)$} g;

output {$R_z(2^k\theta)\ket{\psi_0}$} a;
output {$R_z(2^k\theta)\ket{\psi_1}$} b;
output {$R_z(2^k\theta)\ket{+}$} c;
output {$R_z(2^{k+1}\theta)\ket{\psi_2}$} e;
output {$R_z(2^{k+1}\theta)\ket{+}$} f;

\draw (ct1)+(-1.55,0.75) |- + (-1.32,0.35) ;
\draw (ct1)+(1.32,0.35) -| + (1.52,0.75) ;

\draw (ct2)+(-1.2,0.45) |-  (ct2) ;
\draw (ct2) -| + (1.2,0.45) ;

\end{yquant}
\end{tikzpicture}

\caption{The construction of a 2-layer tower using the compact `CT block'.}
\label{fig:2layer}
\end{figure}
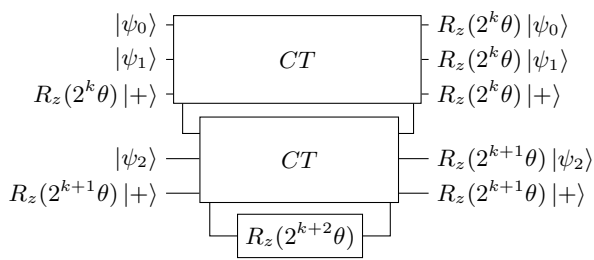

\begin{figure*}
    \centering
\begin{tikzpicture}
\begin{yquant*}
    qubit {$\ket{\psi_0}$} a;
    qubit {$\ket{\psi_1}$} b;
    qubit {$R_z(2^k\theta)\ket{+}$} c;
    qubit {} d;
    qubit {$\ket{\psi_2}$} e; 
    qubit {$R_z(2^{k+1}\theta)\ket{+}$} f;
    qubit {} g;
    discard d,g;

    x c;
    cnot b,c|a;
    zz (b,c);
    [name=ct1l]
    cnot d|a;
    init {} d;

    hspace {2.4cm} f;
    x f;
    cnot e,f|d;
    zz (e,f);
    cnot g|d;
    [name=ct2]
    box {$R_z(2^{k+2}\theta)$} g;
    cnot g|d;
    zz (e,f);
    cnot f|e;
    cnot e,f|d;
    hspace {5.5mm} d;
    x f;
    
    [name=ct1r]
    cnot d|a;
    discard d;
    zz (b,c);        
    cnot c|b;
    cnot b,c|a;
    x c;
    
    output {$R_z(2^k\theta)\ket{\psi_0}$} a;
    output {$R_z(2^k\theta)\ket{\psi_1}$} b;
    output {$R_z(2^k\theta)\ket{+}$} c;
    output {$R_z(2^{k+1}\theta)\ket{\psi_2}$} e;
    output {$R_z(2^{k+1}\theta)\ket{+}$} f;

    \draw (ct1l) + (-0.6,0.5) |- (ct1l);
    \draw (ct1l) -- +(0.6,0);
    \draw (ct1r) -| + (0.6,0.5);
    \draw (ct2)+(-1.78,0.65) |-  (ct2) ;
    \draw (ct2) -| + (1.78,0.65);
\end{yquant*}
\end{tikzpicture}
    \caption{The circuit for the 2-layer tower in \cref{fig:2layer}, note the `V-shape' of the gates, which indicates the circuit is inherently sequential. When used as an `in-circuit' method, this is similar to the technique in \cite{wang2023option}.}
    \label{fig:Vshape}
\end{figure*}
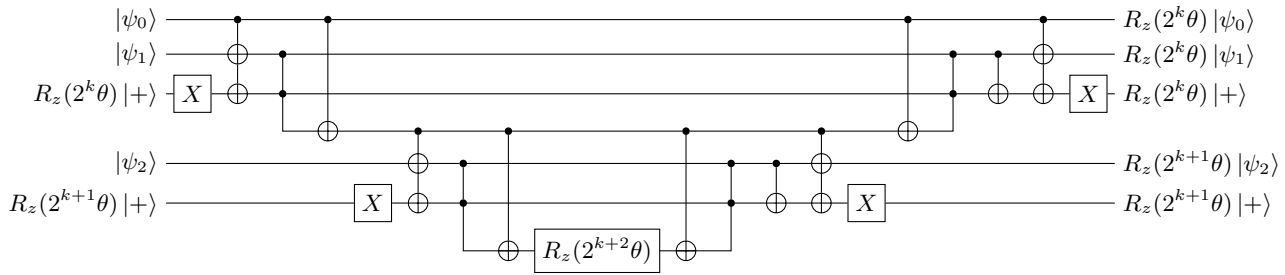

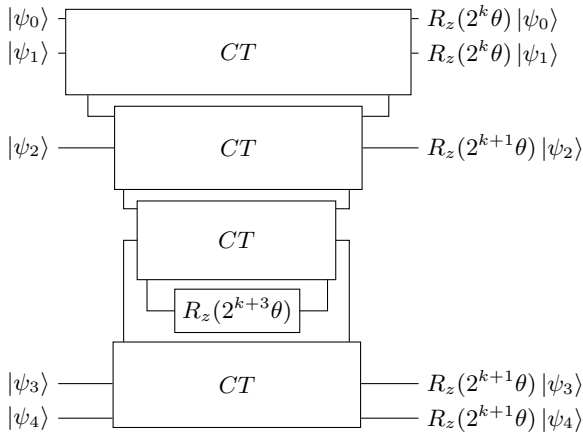
\begin{figure}[h]
    \centering

\begin{tikzpicture}
\begin{yquant*}
qubit {$\ket{\psi_0}$} q1;
qubit {$\ket{\psi_1}$} q2;
nobit q3;
nobit a;
qubit {$\ket{\psi_2}$} b;
nobit c;
nobit d;
nobit e; 
nobit f;
nobit g;
nobit h;
qubit {$\ket{\psi_3}$} i;
qubit {$\ket{\psi_4}$} j;

box {$\qquad\qquad\qquad CT \qquad\qquad\qquad$} (q1,q2,q3);
hspace {6.5mm} b;
[name=ct0]
box {$\qquad\qquad CT \qquad\qquad$} (a,b,c);
hspace {9.5mm} e;
[name=ct1]
box {$\qquad\quad CT \qquad\quad$} (d,e,f);
hspace {14.5mm} g;
[name=ct2]
box {$R_z(2^{k+3}\theta)$} g;
hspace {6.3mm} h;
box {$\qquad\qquad CT \qquad\qquad$} (h,i,j);

output {$R_z(2^k\theta)\ket{\psi_0}$} q1;
output {$R_z(2^k\theta)\ket{\psi_1}$} q2;
output {$R_z(2^{k+1}\theta)\ket{\psi_2}$} b;
output {$R_z(2^{k+1}\theta)\ket{\psi_3}$} i;
output {$R_z(2^{k+1}\theta)\ket{\psi_4}$} j;

\draw (ct0)+(-2,0.7) |- + (-1.65,0.42) ;
\draw (ct0)+(1.65,0.42) -| + (2,0.7) ;

\draw (ct1)+(-1.5,0.67) |- + (-1.32,0.42) ;
\draw (ct1)+(1.32,0.42) -| + (1.5,0.67) ;

\draw (ct2)+(-1.2,0.42) |-  (ct2) ;
\draw (ct2) -| + (1.2,0.42) ;

\draw (ct1)+(-1.5,-1.35) |- (ct1); 
\draw (ct1) -| +(1.5,-1.35); 

\end{yquant*}
\end{tikzpicture}
    \caption{An example of an independent catalyst tower with 3 layers (one way to see this is that the tower is seeded by $R_z(2^{k+3}\theta)$ and the lowest power output is $R_z(2^{k}\theta)$). The catalyst states are omitted for simplicity. For the tower to produce resource states to be used in gate teleportation, set all $\ket{\psi_i}$ to $\ket{+}$ states.}
    \label{fig:trunkleaves}
\end{figure}

The third method attempts to achieve the `best of both worlds', with a depth that is typically below either of the earlier methods and a frugal T-cost comparable to the second. This third, \textit{independent catalyst tower} approach adopts a probabilistic scheme. Instead of preparing the required rotations in place, these towers can be seen as higher-level magic state factories. An example of a 3-layer independent catalyst tower is depicted in \cref{fig:trunkleaves}, where we omit all the catalyst states for simplicity; to generalise it, concatenate one CT block at the top and one at the bottom. The resource states $R(2^i\theta)\ket{+}$ produced by a catalyst tower can then be used to implement a rotation on the target qubit via gate teleportation. However, since the rotation angles can be arbitrary, there is no efficient way to correct for the `wrong' measurement outcome of the gate teleportation directly, so we adopt the repeat-until-success approach with a success probability of 0.5. A `wrong' measurement outcome indicates that the implemented operation was a rotation of the desired angle but in the opposite direction. We now need to teleport a rotation with twice the target angle which can correct the previous failure into a success, again with a probability of 0.5. This process is repeated on failure, and the angle doubles every trial until a measurement `success'. Therefore, the expected measurement depth scales only logarithmically in the number of parallel rotations and thus logarithmically in $n$. This comes with the penalty that we somewhat increase the T-count compared to the in-circuit towers, as we are implementing more rotations on average.

The repeat-until-success process thus involves a doubling of the desired angle, and this is in addition to the binary pattern already occurring within each $\Xi_z$ block; thus the towers' ability to generate multiple rotation resources, related to one-another by powers of two, is well-matched to our task. We design each independent catalyst tower such that a single run of the circuit produces the required species of rotations which closely caps the expected number of rotations required by a group of $n$ qubits (so we only need $S+1$ such towers in principle). However, due to the probabilistic nature of this approach, in practice, it is desirable to have a reservoir of the rotation resource states to account for any statistical fluctuation. As a side note, more generally, the independent catalyst towers are also beneficial in the circuit where multiple rotation gates of the same angle are required (and the circuit is repeatedly applied). One then needs to slightly modify the structure of the tower in \cref{fig:trunkleaves} and implement multiple towers of different layers to match the tower outputs with the distribution of the repeat-until-success scheme.

We note that in \cite{mooney2021cost}, both ways of utilising catalyst circuits, namely the in-circuit and independent mode, are explored using the basic circuit to produce Clifford hierarchy rotations. However, in this work, we achieve our efficiency by composing multi-tier catalyst towers and matching their output to the rotations required by the circuit. As we will see in the next section, the advantage of introducing the composed towers is that, after the initial synthesis of single qubit catalyst states, they achieve a better scaling of the T-count when the circuit is repeated. 

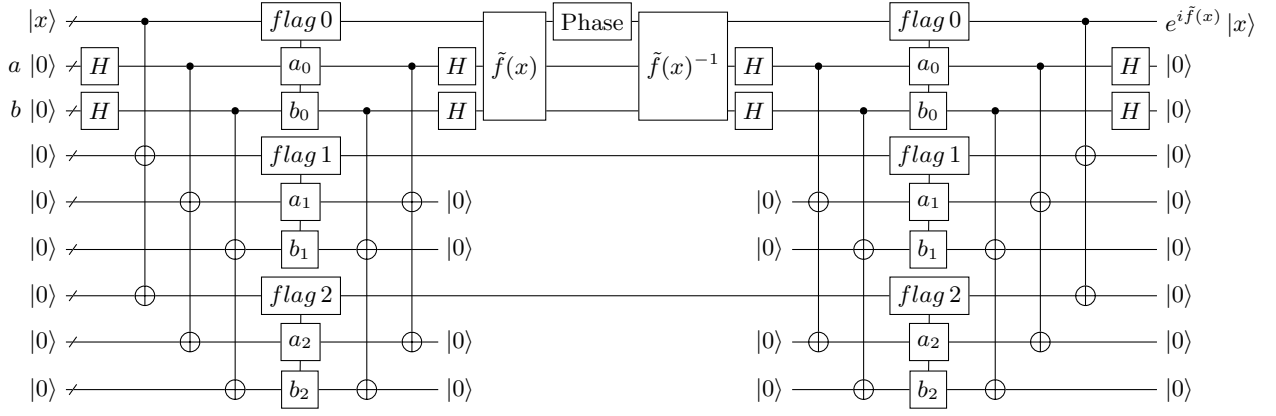
\begin{figure*}
\begin{tikzpicture}
\begin{yquant}
qubit {$\ket{x}$} x;
qubit {$a\, \ket{0}$} a;
qubit {$b\, \ket{0}$} b;
qubit {$\ket{0}$} x1;
qubit {$\ket{0}$} a1;
qubit {$\ket{0}$} b1;
qubit {$\ket{0}$} x2;
qubit {$\ket{0}$} a2;
qubit {$\ket{0}$} b2;
slash - ;
hspace {0.1cm} -;

H a,b;
cnot x2,x1 | x;
cnot a2,a1 | a;
cnot b2,b1 | b;

align x,a,b;
hspace {0.26cm} a,b;
[name=flag0]
box {$flag\,0$} x;
[name=a0]
box {$a_0$} a;
[name=b0]
box {$b_0$} b;

align x1,a1,b1;
hspace {0.26cm} a1,b1;
[name=flag1]
box {$flag\,1$} x1;
[name=a1]
box {$a_1$} a1;
[name=b1]
box {$b_1$} b1;

align x2,a2,b2;
hspace {0.26cm} a2,b2;
[name=flag2]
box {$flag\,2$} x2;
[name=a2]
box {$a_2$} a2;
[name=b2]
box {$b_2$} b2;

cnot b2,b1 | b;
cnot a2,a1 | a;
H a,b;

text {$\ket{0}$} a1,a2,b1,b2;

box {$\tilde f(x)$} (x , a, b);
box {$\text{Phase}$} x;
box {$\tilde f(x)^{-1}$} (x , a, b);

discard a1,a2,b1,b2;

init {~~~~~~~~~~~~~~~~~~~~~~~~~~~~~~~~$\ket{0}$} a1,a2,b1,b2;
H a,b;
cnot a2,a1 | a;
cnot b2,b1 | b;

align x,a,b;
hspace {0.26cm} a,b;
[name=flag0i]
box {$flag\,0$} x;
[name=a0i]
box {$a_0$} a;
[name=b0i]
box {$b_0$} b;

align x1,a1,b1;
hspace {0.26cm} a1,b1;
[name=flag1i]
box {$flag\,1$} x1;
[name=a1i]
box {$a_1$} a1;
[name=b1i]
box {$b_1$} b1;

align x2,a2,b2;
hspace {0.26cm} a2,b2;
[name=flag2i]
box {$flag\,2$} x2;
[name=a2i]
box {$a_2$} a2;
[name=b2i]
box {$b_2$} b2;

cnot b2,b1 | b;
cnot a2,a1 | a;
cnot x2,x1 | x;
H a,b;

output {$e^{i \tilde f(x)} \ket{x}$} x;

output {$\ket{0}$}, a,b,x1,x2,a1,a2,b1,b2;

\end{yquant}
\draw (flag0) -- (a0);
\draw (a0) -- (b0);

\draw (flag1) -- (a1);
\draw (a1) -- (b1);

\draw (flag2) -- (a2);
\draw (a2) -- (b2);

\draw (flag0i) -- (a0i);
\draw (a0i) -- (b0i);

\draw (flag1i) -- (a1i);
\draw (a1i) -- (b1i);

\draw (flag2i) -- (a2i);
\draw (a2i) -- (b2i);

\end{tikzpicture}
\caption{Illustration of the parallelisation of the application of a phase via QROM and linear interpolation as described in \cite{sandersCompilationFaultTolerantQuantum2020}, using only one phase application to save resources. Here $a_i, b_i$ indicate applying circuits that apply a string of Pauli $Z$s, loading in the binary value of the slope and intercept for that section when condition on flag $i$ being set. The four bundles of ancilla qubits used for loading the $a_i, b_i$ are switched off after fan-in; these qubits can be repurposed during the $\tilde{f}(x)$ and Phase computations. They are then re-initialised to be used in the second layer of flag computations. Note that as with Fig. \ref{fig:mainCirc}, this circuit can be optimised using the methods outlined in Appendix \ref{appx:logicaland}.}
\label{fig:qrom_parallel}
\end{figure*}

We compare the aforementioned piecewise function implementation to the method based on a parallelisation of linear interpolation via QROM as described in \cite{sandersCompilationFaultTolerantQuantum2020}. QROM is used to load constants in the linear interpolation of a function $\tilde f(x) = a_i x +b_i$, where the slope and intercept $a_i,b_i$ vary with $x$. The calculation of $\ket{\tilde f(x)}$ can then be performed with a single multiplication and addition, with the phase applied via addition into a phase gradient state. We discuss in more detail about the choices we make for the arithmetic components in \cref{appx:qrom_comments} and provide a circuit demonstrating the parallelised QROM approach in \cref{fig:qrom_parallel}.

\begin{table*}[t]
    \centering
    \begin{tabular}{p{3cm}|p{4.25cm}|p{4.05cm}|c|c|c}
         & T-count (for $r$ rounds) & Meas. depth (per round) & A. Pricing & B. Coulomb & C. QD \\
         \hline
         Gate synthesis& $r(S{+}1)(n{+}1)rot_T {+} 8Sr(l{-}1)$  & $rot_T {+} 2\lceil \log_2 l \rceil$ & 16536/32 & 3582/27 & 2214/23 \\
         \hline
          Gate synthesis with injection& $2r(S{+}1)(n{+}1)rot_T {+} 8Sr(l{-}1)$  & $\lceil\log_2 (2.5(S{+}1)(n{+}1){+}1.5)\rceil\newline{+} 2\lceil \log_2 l \rceil$ & 31633/17 & 6852/13 & 4116/12 \\
         \hline
         In-circuit towers& $(S{+}1) \{[rot_T{\cdot}(n{+}1){+}4n]\newline{+}(r{-}1)(rot_T{+}4n){+}r\cdot rot_T\}\newline {+} 8Sr(l{-}1)$ & $rot_T {+} 2n {+} 2\lceil \log_2 l \rceil$ & 5618/62 & 1476/45 & 1191/35 \\
         \hline
         Independent towers& $(S{+}1) \{[rot_T\cdot(2n{+}3) {+} 8n{+}4]\newline{+}(r{-}1)(2rot_T{+}8n{+}4)\}\newline{+} 8Sr(l-1)$ & $\lceil\log_2 (2.5(S{+}1)(n{+}1){+}1.5)\rceil\newline{+} 2\lceil \log_2 l \rceil$ & 8061/17 & 2042/13 & 1583/12 \\
         \hline
        Linear interpolation with QROM & $n \cdot R_T + r(14 \cdot 3^{\lceil \log_2(n) \rceil} + 66n)\newline{+} 8Sr(l-1)$ & $9\lceil \log_2(n) \rceil +1+ 2\lceil \log_2 l \rceil$ & 3566/43 & 2040/41 & 1086/32 \\
         
    \end{tabular}
    \caption{The T-count and measurement depth of the 4 options discussed in \cref{sec:options}. For independent towers method, we assume the depth for resource state production does not dominate (see discussions in \cref{sec:cost}). The expressions are evaluated for the three examples in \cref{sec:examples}, i.e. A. option pricing (accuracy, $\epsilon_{circ}=10^{-3}$), B. Coulomb interaction ($\epsilon_{circ}=10^{-3}$), C. quantum dots confining potential ($\epsilon_{circ}=10^{-2}$). The numerical results are in the form T-count per round/measurement depth per round. We note that, in these examples which have relatively small $n$ values, the $O(n^{\log_2(3)})$ T-count of the linear QROM method outperforms some of the other methods that also depend on $S$ (see \cref{subsec:Coulomb}). We also note that the QROM method is the only method whose $T$-count or depth of the central component (the parts excluding the flags) does not increase with $S$, making it the preferred choice for sufficiently complex functions.}
    \label{tab:expressions}
\end{table*}

\section{Cost analysis} \label{sec:cost}
We use measurement depth as the metric for circuit depth and let a single qubit measurement in $Z$ or $X$ basis have depth 1. We use this metric instead of T-depth to account for teleporting the rotation resource states and also the uncomputation of the AND gate, which have the same delay as the conventional T-state teleportation circuit on fault-tolerant level (assuming we are not limited by magic state production rate). Therefore, T-state injection (the T-depth), rotation resource state injection and the uncomputation of AND all have measurement depth 1.

Let $rot_T$ be the T-count of a rotation gate synthesized using Clifford+T. For a given accuracy $\epsilon$, the optimal T-count for deterministic and ancilla-free approximation of a rotation gate is $3\log_2{(1/\epsilon)}$ \cite{RossSelinger2014}. Given that we are in a regime that allows for an arbitrary number of ancilla, we can use the fallback, or projective rotation, approximation method \cite{BocharovRoettelerSvore2014}, which reduces the T-count of a rotation \cite{kliuchnikov2023shorter} to
\begin{equation}
rot_T \approx 1.03\log_2{(1/\epsilon)+5.75}.
\end{equation}
 Using probabilistic mixing methods would further reduce this cost by a factor of 2 \cite{Campbell2017, Hastings2017, kliuchnikov2023shorter}. To implement a circuit with $k$ rotations to precision $\epsilon_{circ}$, take the error per rotation as $\epsilon = \epsilon_{circ}/k$ when calculating $rot_T$. Since we are using Clifford+T, $rot_T$ is also the T-depth and measurement depth.
 
We also note that the following expressions and calculations use one  $rot_T$ parameter for all the rotations, this is enough for illustration here. However, for a more detailed analysis, one might want to use a different $rot_T$ for catalyst states than other rotations (the choice could be highly empirical) to further improve the performance. 

For canonical gate synthesis, the T-count of the circuit is
\begin{equation}
    (S+1)(n+1)rot_T + 8S(l-1) 
\end{equation}

\noindent where $S$ is the number of sections, and $n$ is the number of qubits in a register. The first term is the contribution from the rotation towers, and the last term is due to the flag blocks. For the flag blocks we need $2S$ $l$-controlled Toffoli ($l=\lceil \log_2{S} \rceil$) which can be constructed from $l-1$ regular Toffoli. Using the construction in \cite{PhysRevA.87.022328}, a Toffoli can be constructed with a T-count of 4 and injected into the circuit with depth 1. Combined with the binary tree structure in \cite{he2017decompositions}, a $l$-controlled Toffoli has a T-count of $4l-4$ and measurement depth $\lceil \log_2 l \rceil$.

The measurement depth of the circuit is
\begin{equation} 
rot_T + 2\lceil \log_2 l \rceil.
\end{equation}
This method can alternatively be coupled with state injection to reduce circuit depth, at the cost of increased T count. The required rotations are synthesised as resource states then injected in parallel with a success probability of 0.5. We expect to require two such resource states per rotation angle, hence the increase in T count to \begin{equation}2(S{+}1)(n{+}1)rot_T {+} 8S(l{-}1).\end{equation} The measurement depth is reduced to \begin{equation} \lceil\log_2 (2.5(S+1)(n+1)+1.5)\rceil+ 2\lceil \log_2 l \rceil,\end{equation} which we explain further at the end of this section.

Meanwhile for the in-circuit catalyst towers, the T-count of the circuit for $r$ rounds of repetition (that is, $r$ invocations of the phase oracle) is
\begin{equation}
\begin{split}
(S{+}1) \{[rot_T\!\cdot\!(n+1)+4n&]{+}(r{-}1)(rot_T{+}4n) {+}r\!\cdot\! rot_T\} \\ &+ 8Sr(l-1)
\end{split}
\end{equation}
and the measurement depth of the circuit is
\begin{equation} 
rot_T + 2n + 2\lceil \log_2 l \rceil
\end{equation} 

\noindent where the first two terms are the contribution from the towers. Each $n$-layer tower also needs $2n+1$ extra ancillary qubits, which is $(S+1)(2n+1)$ ancillary qubits in total (this does not include the ancillae used by the fallback synthesis method).

Finally for the independent catalyst towers, the T-count of the circuit for $r$ rounds is
\begin{equation} 
\begin{split}
&(S{+}1) \{[rot_T\!\cdot\!(2n{+}1) {+} 8n]{+}(r{-}1)[rot_T{+}8n]\} \\ &+(S{+}1) \{(rot_T\!\cdot\!2 {+} 4){+}(r{-}1)(rot_T{+}4)\}+ 8Sr(l-1)
\end{split}
\end{equation}

\noindent where the first term is the contribution of the independent towers associated with the large rotation blocks, and the second term is from the 1-layer independent towers that deal with the single qubit $\gamma_i$ rotations. This way of doing the $\gamma_i$ rotations is more costly in terms of T-count but it reduces the depth of the circuit.

The expected measurement depth of the circuit is
\begin{equation} \label{eq:inde_depth}
\lceil\log_2 (2.5(S+1)(n+1)+1.5)\rceil+ 2\lceil \log_2 l \rceil.
\end{equation} 

Intuitively, when applying $M$ parallel rotations using the repeat-until-success scheme, $M/2$ rotations are expected to fail, which requires a further round of rotations with $M/4$ expected failures, etc. The contribution to depth of this process is captured by the first term of \cref{eq:inde_depth}, which is logarithmic in the number of parallel rotations. We provide a derivation of this expected depth in \cref{appx:depth}, with the expression above being a numerical fitting that is within 0.04 of the true expectation for all $m\le10^6$, where $m$ is the number of parallel rotations. For completeness, an $n$-layer ($n\geq 2$) independent tower uses $6n-5$ qubits and has a measurement depth of $rot_T + 2n$. Note that this depth of the independent towers (which act like magic state factories) scales worse than the depth of the phase oracle circuit that consumes the resource states (\cref{eq:inde_depth}). Therefore, in the cases where the phase oracle dominates the depth of the full circuit being repeated, one should use $rot_T + 2n$ as the phase oracle depth. As the focus of this work is minimising the depth, we note a relevant mitigation: one can break down an $n$-layer independent tower into several shorter towers, e.g. $n-1$ 1-layer towers (in \cref{fig:1layer}) with constant depth, at the cost of consuming more T states. The modular nature of the CT blocks makes this reconstruction simple.

However, in many applications the phase oracle is used as a subroutine of a much longer circuit, and such cases are suitable for the independent towers paradigm. For example, in the option pricing application \cite{chakrabarti2021threshold} in the next section, each repetition for amplitude estimation involves preparing Gaussian states, followed by quantum arithmetic, and then applying the phase oracle. This allows enough time for resource state production to take place between the instances of the oracle. 

The above expressions are summarised in \cref{tab:expressions}. From there it is clear that for repetition count $r=1$, neither catalyst methods give any advantage over direct gate synthesis (without injection). However, only a modest number of repetitions is required to reach the break-even point for T-count; for the in-circuit tower it needs

\begin{equation}
    r > \left(1-\frac{1}{n}-\frac{4}{rot_T}\right)^{-1},
\end{equation}
while for the independent towers it requires more
\begin{equation}
    r > \left(1-\frac{n+2}{2n+1}-\frac{4}{rot_T}\right)^{-1}
\end{equation}

For the option pricing application in the next section, where $S=36, n=15, \epsilon_{circ}=10^{-3}, r=200, l=6$, the in-circuit and independent towers need 2 and 4 rounds to bring advantage respectively. Note that this analysis is at the fault-tolerant level. For a more complete calculation of these thresholds, one needs to consider the spacetime volume of the towers and the T-factories at the quantum error correcting code level.

As a rule of thumb for choosing between the towers when $r$ is larger than the threshold value, if the goal is to minimise the T-count then one should use the in-circuit towers. If the goal is to minimise the circuit depth (even when $r$ is small), one might want to use the independent towers. As for the QROM method, it can be competitive in terms of T-count when $n$ is small, but one should note that the scaling is worse than linear (see discussions in \cref{subsec:Coulomb}).

Finally, we used T-count and T-depth above as our metrics for the resource costs of our oracles. However, we note that recent development in magic state production \cite{gidney2024magic} indicates that T gates can be generated at a cost comparable to a lattice surgery CNOT. Therefore, it might be worth noting that $O(S\cdot n)$ CNOT gates are also required for the circuit.

\section{Examples} \label{sec:examples}
In this section, we present explicit piecewise solutions and consequent resource counts for contexts involving
\begin{enumerate}
\item[A.] payoff functions for option pricing,
\item[B.] the Coulomb interaction, and
\item[C.] a `double dot' confining potential.
\end{enumerate}

\subsection{Payoff function for option pricing}

As our first example, we use part of the payoff function for option pricing that appears in \cite{stamatopoulos2024derivative}, in which the authors use quantum signal processing (QSP) to replace the quantum arithmetic in \cite{chakrabarti2021threshold}. For this example, we need the variant which implements $(f(x)\ket{0}+\sqrt{1-f(x)^2}\ket{1})\ket{x}$
(where we change the target definition slightly to match the formalism in Ref.~\cite{stamatopoulos2024derivative} more closely), as this is required by the subsequent iterative quantum amplitude estimation \cite{grinko2021iterative} step in their algorithm. For their use case, the circuit can be repeatedly applied up to $\sim10^2$ iterations \cite{labib2024quantum} and each iteration calls the oracle twice.

The target function to encode is

\begin{equation} \label{eq:target_derivatives}
    f(x) = \sqrt{1-(K_T - e^{x \cdot 2^p} e^{-s})},
\end{equation}
with the parameters $K_T = 1$, $p = 5$, and $s=2^5$. This closely resembles Eq.~(46) in \cite{stamatopoulos2024derivative} with the parameters of Fig.~4a of the same reference. The target here differs from the one used in Ref.~\cite{stamatopoulos2024derivative} by the square of $x$ in the exponent, which is required for QSP in~\cite{stamatopoulos2024derivative} to work, but we can omit it here.

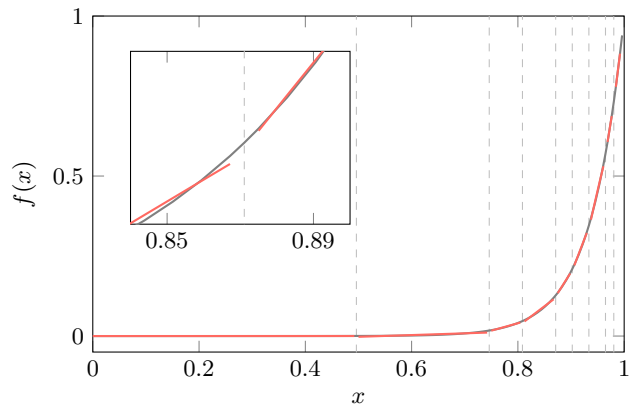
\begin{figure}
    \centering
    \begin{tikzpicture}

\definecolor{pastelred}{rgb}{1.0, 0.41, 0.38}

\begin{axis}[
    height=.7\columnwidth,
    width=\columnwidth,
    xmin=0, xmax=1,
    ymin=-.05, ymax=1,
    unbounded coords=jump,
    xlabel={$x$}, ylabel={$f(x)$}
    ]
    \addplot[mark=none, color=lightgray, dashed, thin] table {data/payoff_borders.dat};
    \addplot[mark=none, color=gray, thick] table[x=x, y=targ] {data/payoff_data.dat};
    \addplot[mark=none, color=pastelred, thick] table[x=x, y=fit] {data/payoff_data.dat};
\end{axis}

\begin{axis}[
    xshift=.5cm, yshift=1.7cm, width=4.5cm,
    xmin=.84, xmax=.9,
    ymin=.08, ymax=.18,
    ytick=\empty, xtick={.85, .89},
    unbounded coords=jump,
    axis background/.style={fill=white}
    ]
    \addplot[mark=none, color=lightgray, dashed, thin] table {data/payoff_borders.dat};
    \addplot[mark=none, color=gray, thick] table[x=x, y=targ] {data/payoff_data.dat};
    \addplot[mark=none, color=pastelred, thick] table[x=x, y=fit] {data/payoff_data.dat};
\end{axis}

\end{tikzpicture}
    \caption{Sectioning of the function in \cref{eq:target_derivatives} with the parameters mentioned in the text thereafter. The target function is in grey, and the linear segments used for approximating the target are in red. Because of the similarity of the curves, the inset shows a close-up of segment edges. Dashed vertical lines show the edges of segments. For better visualisation, the depicted segmentation uses a qubit count of 7 and a somewhat large tolerance between the target and segmented functions of $\max_x \lvert f_\mathrm{target}(x) - f_\mathrm{segmented}(x)\rvert \leq 10^{-2}$. More realistic values for these hyperparameters and their implications are discussed in the main text. Note that the possible $x$-values are not continuous, but given by the discretisation representable by the used number of qubits. Thus the gaps in the linear approximation signify that there are no data points (quantum states) between the last value of the previous and the first value of the next linear section.}
    \label{fig:linear_derivatives}
\end{figure}

\Cref{fig:linear_derivatives} shows an intuitive visualisation of the piecewise approximation. The depicted segmentation uses a quantum state of only 7 qubits and a high tolerance threshold of
\begin{equation}
    \max_x \lvert f_\mathrm{target}(x) - f_\mathrm{segmented}(x)\rvert \leq 10^{-2}
\end{equation}
as mentioned in the Figure caption. This results in a total of 9 segments. However, practically more relevant parameters would be, for example in this case, 15 qubits and a tolerance of only $10^{-3}$, which then results in 36 segments. From \cite{chakrabarti2021threshold, stamatopoulos2024derivative}, we use $S=36, n=15, \epsilon_{circ}=10^{-3}, r=2\times10^2, l=6$ for the expressions in \cref{sec:cost}, the cost of each method is shown in \cref{tab:expressions}. We can see that the in-circuit towers can reduce the T-count per round by a factor of 2.9 but have a factor of 1.9 increase in depth, in comparison to direct gate synthesis. The independent towers achieve a factor of 2.1 reduction in T-count and a factor of 1.9 reduction in depth. 

The width-depth trade-off is also favourable in this case. The quantum signal processing method in \cite{stamatopoulos2024derivative} results in a qubit count $\sim 30$ and a T-depth of $\sim 5700$. In comparison, the deterministic in-circuit towers, for example, need $(S+1)(3n+2) \sim 1700$ qubits with measurement depth 62. So our approach has a factor of $\sim57$ increase in width but achieves a factor of $\sim 92$ reduction in depth. This improvement in depth can be vital for applications like derivative pricing where the time, usually on the scale of seconds, is critical for delivering quantum advantage. It is also worth noting that the prior quantum arithmetic step required by the option pricing algorithm~\cite{stamatopoulos2024derivative} is much wider compared to the phase oracle, so the width-depth trade-off here does not actually incur new qubit overhead. In essence, the qubits are available anyway and thus {\it may as well} be put to use in the fashion we describe.

Note that on a larger scope, since the phase oracle is usually just a subroutine of an algorithm, reusing the catalyst (so they exist throughout the algorithm) states can cause the width of other parts of the circuit to increase.

\subsection{Coulomb interaction}\label{subsec:Coulomb}
The next illustration we show is that of approximating a Coulomb potential, as is required by many algorithms in the context of materials science or chemistry. Prominent examples are ones leveraging grid-based methods to encode the state of (electronic) systems directly in the amplitudes of a register~\cite{hans_grid_based_2023, jnane_ab_initio_2024}. 
Such methods involve using quantum registers to represent particles on a discretised grid and often use the Split-Operator Quantum Fourier transform (SO-QFT) method to drive the dynamics. When combining with Trotterization, the method can apply time evolution by repeatedly transforming between real space and momentum space. The potential terms of the first-quantised Hamiltonian can be approximated as diagonal in real space (and the kinetic terms are diagonal in momentum space). Our methods for effecting a phase $\exp(iV(\mathbf{r}))$ onto the amplitudes is a natural fit for this simulation problem with $V(\mathbf{r})$ being the Coulomb potential. It has been shown in \cite{babbush2023quantum} that such simulation algorithms have superquadratic speedup (quartic in certain regimes) over classical methods, while the effort to establish exponential speedup remains an active area of research \cite{dalzell2023quantum}.

\begin{figure}
    \centering
    \begin{tikzpicture}

\definecolor{pastelred}{rgb}{1.0, 0.41, 0.38}

\begin{axis}[
    height=.7\columnwidth,
    width=.9\columnwidth,
    axis y line*=left,
    xmin=0, xmax=1,
    ymin=1e-3, ymax=1,
    ymode=log,
    unbounded coords=jump,
    xlabel={$x$}, ylabel={$f(x)$},
    axis on top,
    ]
    \addplot[mark=none, color=darkgray!70, thick] table[x=x, y=target] {data/coulomb_data.dat};
    \addplot[mark=none, color=pastelred, thick] table[x=x, y=fit] {data/coulomb_data.dat};
\end{axis}

\begin{axis}[
  height=.7\columnwidth,
  width=.9\columnwidth,
  axis y line*=right,
  axis x line=none,
  ymin=1e-5, ymax=1e-3,
  xmin=0, xmax=1,
  ymode=log,
  ylabel={Error},
  scaled ticks=false,
  axis on top,
]
    \addplot[mark=none, color=black!80, semithick] table[x=x, y=err] {data/coulomb_data.dat};
\end{axis}

\end{tikzpicture}
    \caption{Piecewise linear approximation of a (singularity-lifted) 1D Coulomb potential using 9 qubits to represent $x$ and a tolerance of $10^{-3}$. The target is the grey $x^{-1}$ curve with the (interrupted, as in \cref{fig:linear_derivatives}) approximation in red on top. The black line shows the error across $x$. This approximation requires 13 sections in the piecewise representation to acquire the desired accuracy.}
    \label{fig:coulomb_potential_approx}
\end{figure}
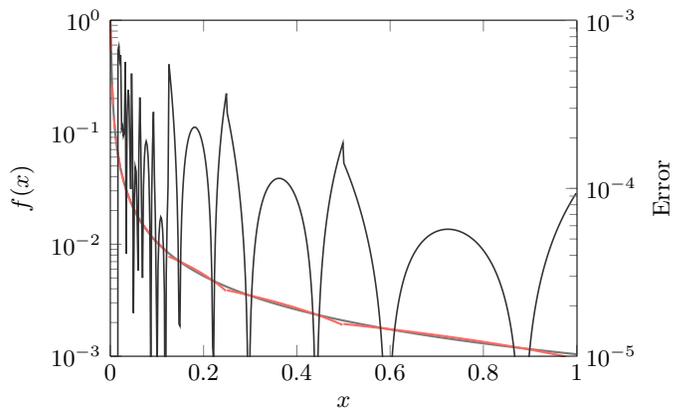

The one-dimensional example shown in \cref{fig:coulomb_potential_approx} with $f(x) = V(x)$ uses $9$ qubits and tolerates a maximum deviation from the target function by $10^{-3}$. The singularity at $x=0$ is replaced by $f(0) = 2^9 \cdot 30 / 16$, as suggested by Ref~\cite{jnane_ab_initio_2024}. For comparability with the other examples, the function is also scaled down to have a maximum value of 1. This example requires 13 sections to approximate to the desired accuracy. Based on \cite{hans_grid_based_2023}, we might need at least $r\sim500$ to give a useful simulation (possibly far more). Assuming  $S = 13, n = 9$ and $\epsilon_{circ}=10^{-3}$ for convenience, the cost of each method is computed in \cref{tab:expressions}.  It might appear that the register size $n=9$ suggests a small problem size. Yet the grid-based method works by allocating relatively small registers of this kind to each particle (or particle dimension), while the total number of modelled particles can be large. Importantly, we only need to apply two-body Coulomb interactions to these modest sized registers, albeit this should be performed in parallel over multiple registers. Indeed this also applies to the quantum dot example in the next section. It is worth noting that the tolerable level of imperfection $\epsilon_{circ}$ is highly dependent on the exact use case and will be determined in combination with other parameters like time step. We use $\epsilon_{circ}=10^{-3}$ here as an example, since the error from other sources e.g. the Trotterisation formula may be expected to be of this order \cite{ikeda2024measuring}.

\subsection{Double quantum dot}
\begin{figure}
    \centering
    \begin{tikzpicture}

\definecolor{pastelred}{rgb}{1.0, 0.41, 0.38}

\begin{axis}[
    height=.7\columnwidth,
    width=.9\columnwidth,
    axis y line*=left,
    xmin=0, xmax=1,
    ymin=0, ymax=1,
    unbounded coords=jump,
    xlabel={$x$}, ylabel={$f(x)$},
    axis on top
    ]
    \fill[gray!10] (0,0) rectangle (.2,1);
    \fill[gray!10] (.8,0) rectangle (1,1);
    \addplot[mark=none, color=darkgray!70, thick] table[x=x, y=target] {data/dot_data.dat};
    \addplot[mark=none, color=pastelred, thick] table[x=x, y=fit] {data/dot_data.dat};
\end{axis}

\begin{axis}[
  height=.7\columnwidth,
  width=.9\columnwidth,
  axis y line*=right,
  axis x line=none,
  ymin=6e-5, ymax=2e0,
  xmin=0, xmax=1,
  ymode=log,
  ylabel={Error},
  scaled ticks=false,
  axis on top
]
    \addplot[mark=none, color=black!80, semithick] table[x=x, y=err] {data/dot_data.dat};
\end{axis}

\end{tikzpicture}
    \caption{Sectioning of a 1D double quantum dot potential. For this type of function, we can reduce the number of segments by choosing different tolerances in different zones. The grey areas can tolerate more errors as long as they provide enough confinement as explained in the main text. For illustration purposes, we chose tolerances equal to $10^{-3}$ and $10^{-1}$ in the non-shaded and shaded areas respectively. The error between the curves is plotted in black.}
    \label{fig:dqd_potential_approx}
\end{figure}
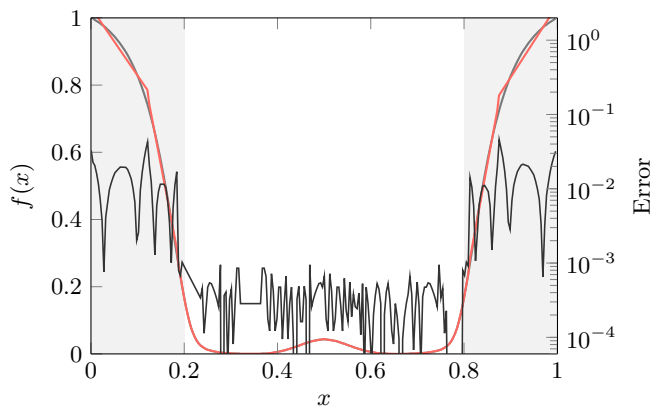

As our final example, we examine a double quantum dot potential, which appears in the study of quantum dot structures. Typically, they are explored using classical techniques \cite{cifuentes_2023, shehata_2023, jnane_ab_initio_2024} for the development of semiconductor-based quantum computers. However, future quantum devices could help simulate more general nanostructures using grid-based methods \cite{hans_grid_based_2023, jnane_ab_initio_2024}. Within this formalism, as in the previous example, one will have to apply functions of the type $\exp(iV(\mathbf{r}))$ to a given quantum state, with $V(\mathbf{r})$ meaning a double dot potential as in \cref{fig:dqd_potential_approx}. 

Here, we focus on a one-dimensional double quantum dot potential $V(x)$ and a quantum state of 6 qubits, but this approach can be extended to multiple dimensions supposing that the potential is separable (or nearly so). Such a potential is generated by a Poisson solver applied to a realistic device. 

In \cref{fig:dqd_potential_approx}, we plot a representation of the piecewise approximation. Unlike the other functions studied above, it is not necessary to accurately represent the potential over the entire range of $x$. As long as the potential in the shaded region provides adequate confinement, important properties of the double quantum dot structure such as the wavefunction of the first two eigenstates (used to define the qubit states and extract the exchange coupling \cite{jnane_ab_initio_2024}), are approximately preserved. This allows us to significantly reduce the number of segments needed. 

For instance, using a tolerance of $10^{-3}$ over both zones results in a $55$-segment approximation, achieving an infidelity approximately equal to $\mathcal{I}_g \approx 10^{-8}$ defined as $\mathcal{I}_g = 1-|\braket{g| g_{pw}}|$ with $\ket{g}$, the exact ground state (as computed in \cite{jnane_ab_initio_2024}) and the approximate one $\ket{g_{pw}}$ obtained using the piecewise approximation. By applying the same tolerance within the non-shaded region and a tolerance of $10^{-1}$ in the shaded region, we can reduce the number of segments to $31$ while obtaining an infidelity $\mathcal{I}_g \approx 10^{-6}$ which is enough for our purposes. This is because the wavefunctions vanish in the shaded region, making them robust to modifications of the potential in that zone. It is worth noting that we obtain similar results for the infidelity $\mathcal{I}_e$ computed between the excited states. 

Finally, one can further reduce the number of segments by using a tolerance of $10^{-2}$ and $10^{-1}$ in the non-shaded and shaded areas respectively. This choice yields a $13$-segment approximation while maintaining an infidelity $\mathcal{I}_g \approx 10^{-6}$.

We estimate a useful simulation would require at least $ 5 \times 10^{5}$ time steps \cite{jnane_ab_initio_2024}. Using  $ r = 5 \times 10^{5}$ (for the potential part of Hamiltonian only), $S = 13, n = 6$ and $\epsilon_{circ}=10^{-2}$ for convenience. This is a suitable set of parameters for the 1D scenario, albeit in practice we may need to apply $V(x,y,z)$. The cost of each method is shown in \cref{tab:expressions}. For this large number of repetitions, we are in the asymptotic regime that, e.g. the in-circuit catalyst towers can apply $n$ rotations ($n+1$ rotations strictly speaking) by consuming only $rot_T+4n$ T states (one seed rotation plus $n$ logical AND), which would cost $n\cdot rot_T$ using the canonical gate synthesis.

We note that for these parameters, it seems that the independent towers outperform the QROM approach in depth while the QROM approach is more advantageous in terms of T-count. When moving to larger $n$, the $O(n^{\log_2(3)})$ term in T-count will dominate over the linear term when using multiplication methods that are depth-competitive with the independent towers, refer to \cref{sec:prior} for discussion.

\section{Conclusion} \label{sec:conclusion}

In this work, we study the task of applying a phase $\exp(i\,f(x))$ to a computational basis state $\ket{x}$, and the closely-related task of applying a rotation dependent on $f(x)$. These tasks are well-studied in the literature; our aim here is minimising the depth (i.e. time cost) of the process. We consider a piecewise approximation to $f(x)$ and fully parallelise the process so that its rotation-depth is unity. 

Noting that in many applications, such `phase oracles' are applied numerous times within an algorithm, we explore a means to reduce the T-gate resource cost by use of `catalyst towers', bespoke hierarchical structures based on the catalyst circuits from Ref.\, \cite{gidney2019efficient}. We find that the catalyst towers align very naturally with the requirements of the parallelised piecewise circuit, and can considerably reduce the T-count for the rotation gates if the circuit is repeatedly applied. We also give explicit expressions for T-count and measurement depth of the different methods and compare their behaviour in three examples.

As for future research, it would be desirable to have a more detailed cost analysis of towers at the quantum error-correcting code level. As mentioned at the end of \cref{sec:cost}, one may want to compare the spacetime volume of the T factories saved due to the reduction in T-count with that of the towers added. Moreover, one must also account for the storage cost of the catalyst states, which could manifest as an increase in code distance.

\section{Acknowledgements}

The numerical modelling involved in this study made
use of the Quantum Exact Simulation Toolkit (QuEST) \cite{jones2019quest} via the QuESTlink\,\cite{jones2020questlink} frontend. We are grateful to those who have contributed to all of these valuable tools.
Aspects of this study were supported by
the EPSRC projects Robust and Reliable Quantum Computing (RoaRQ, EP/W032635/1), Software Enabling Early Quantum Advantage (SEEQA, EP/Y004655/1), the EPSRC QCS Hub (EP/T001062/1) and the UKRI Future Leaders Fellowship (MR/Y015843/1). Richard Meister is partially supported by the EPSRC Grant number EP/W032643/1. Romy Minko and Benjamin Pring were partially supported by the Additional Funding Programme for Mathematical Sciences, delivered by EPSRC (EP/V521917/1) and the Heilbronn Institute for Mathematical Research.

\appendix

\section{Circuitry} \label{appx:logicaland}
The computation and uncomputation of the Logical-AND gate, which cost 4 T states and a measurement.

\begin{tikzpicture}
\begin{yquantgroup}
\registers{
    qubit {} a;
    qubit {} b;
    qubit {} c;
}
\circuit{
    init {$x$} a;
    init {$y$} b;
    discard c;
    
    [name=d1]
    zz (a,b);
    
    output {$x$} a;
    output {$y$} b;

    \draw (d1) |- + (0.35,-1.2); 
    \node at ($(d1) + (0.68,-1.23)$) {$xy$};
}
\equals
\circuit{
    init {$\ket{T}$} c;

    cnot c|a;
    cnot c|b;
    cnot a,b|c;
    box {$T^\dag$} a;
    box {$T^\dag$} b;
    hspace {0.5mm} c;
    box {$T$} c;
    cnot a,b|c;
    h c;
    box {$S$} c;
}

\end{yquantgroup}
\end{tikzpicture}

\begin{figure}[h]   
\hspace*{-4.2cm}
\begin{tikzpicture}
\begin{yquantgroup}
\registers{
    qubit {} a;
    qubit {} b;
    qubit {} c;
}
\circuit{
    init {$x$} a;
    init {$y$} b;
    discard c;
    
    [name=d1]
    zz (a,b);
    
    output {$x$} a;
    output {$y$} b;

    \draw (d1) |- + (-0.35,-1.1); 
    \node at ($(d1) + (-0.68,-1.1)$) {$xy$};
}
\equals
\circuit{

    h c;
    measure c;
    box {$Z$} b|a,c;
    discard c;    
}

\end{yquantgroup}
\end{tikzpicture}
\caption{The circuit diagram for the `corner' notation used in the main text, reproducing Figure 3 of \cite{Gidney2018halvingcostof}. }
\label{fig:AND}
\end{figure}
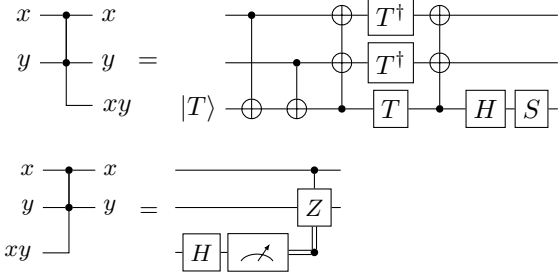

We make heavy use of both fan-out and fan-in via CNOTs in this paper, hence it is worthwhile remarking upon implementation details for these primitives. 

For fanout to enable us to have parallel access to $k$ copies of the $n$-qubit computational basis state $\ket{x_1 \ldots x_n}$, we want to implement $n$ parallel copies of the map $\ket{x_i}\ket{0^{k-1}} \mapsto \ket{x_i}\ket{x_i^{k-1}}$ where $x_i \in \{0,1\}$. Whilst certain technologies allow for multi-target CNOT gates as a primitive \cite{litinski2018lattice}, the real time long-range routing might be a problem at the quantum error correcting code level. Working with an abstract model of quantum computation for now, alternatively, by using one Hadamard gate and $k-1$ CNOT gates, we can prepare a $k$-qubit GHZ state $\frac{1}{\sqrt{2}}(\ket{0}\ket{0^{k-1}} + \ket{1}\ket{1^{k-1}})$ and transport these qubits to where they will be required ahead of time. This state can then implement the required fanout using a single CNOT gate from our control to the first qubit of the GHZ state to obtain $\ket{x}\otimes \frac{1}{\sqrt{2}}(\ket{0}\ket{0^{k-1}} + \ket{1}\ket{1^{k-1}}) \mapsto \ket{x}\frac{1}{\sqrt{2}}(\ket{x}\ket{0^{k-1}} + \ket{\bar{x}}\ket{1^{k-1}})$ followed by a Z-basis measurement of the second qubit of the resulting state which gives us either $0$, in which case we have $\ket{x}\ket{x^{k-1}}$ or $1$, in which case we have $\ket{x}\ket{\bar{x}^{k-1}}$ and a Pauli correction of $X^{\otimes k-1}$ can be made to obtain the desired outcome.

For fan-in, where we wish to accomplish the map $\ket{x}\ket{x^{k-1}} \mapsto \ket{x}\ket{0^{k-1}}$, this can be accomplished via multi-CNOT gates, but it is easier to replace multiple-qubit operations which may require routing and additional circuitry with measurement-based uncomputation. Recall that if we have the two-qubit state $\ket{x}\ket{x}$ and perform a destructive X-basis measurement on the second qubit, we obtain the state $\ket{x}$ if we measure $0$ and $(-1)^x\ket{x}$ if we measure $1$. We can therefore perform $k-1$ such parallel X-basis measurements and perform the Pauli-Z correction conditioned upon the joint parity of these measurements.

One further trade-off is possible with regards to the T-count required for computing the flags in \cref{fig:mainCirc} and \cref{fig:qrom_parallel}, which do not illustrate the ancilla required to implement the flags with minimal depth. We recall these flags will be an $l$-controlled Toffoli, for which the fastest method of implementing these functions is to structure them as binary tree using $l-1$ quantum AND gates ($\ket{a,b}\ket{0} \mapsto \ket{a,b}\ket{a\cdot b}$) which can be implemented $4$ T gates or a CCZ state. As soon the the final such AND gate is completed, the rotation stage can begin and if we choose to keep these intermediate computations in memory, the second layer of flag computations can be completed in the same depth but without using any T gates via measurement-based uncomputation. The utility of whether this particular optimisation is worthwhile is dependent upon whether the space could be used in a more efficient manner (whether as ancilla for other unitaries or for repurposing for magic state distillation, etc.) and if the rotation layer is implemented in a fast manner, this may be worthwhile.

\section{Heuristic for sectioning functions}\label{appx:heuristic}
For the examples given in the main text we use a simple heuristic to divide the target function into (quasi-)linear segments. As we shall see, this method results in the particularly simple `flag' blocks of \cref{fig:mainCirc} mentioned in \cref{sec:parallelising}, which is a simple $l$-controlled Toffoli.

The details of the used heuristic differ slightly depending on whether a phase should be applied to a state, i.e. $\ket{x} \mapsto \exp{i f(x)} \ket{x}$, or whether an amplitude state should be prepared as in $\ket{x}\ket{0} \mapsto \ket{x}(f(x) \ket{0} + \sqrt{1 - f(x)^2} \ket{1})$. However, the basic principle remains the same. We will discuss the former case first, followed by a comment on the required changes for the latter case.

The task is to section the target function $f(x)$ into intervals $\Delta x_j$, where within the interval $j$, $A_j + B_j x \approx f(x)$, with some parameters $A_j$ and $B_j$ and $x\in\Delta x_j$. The following bisection method accomplishes this.
\begin{itemize}
    \item Within some interval $\Delta x_j$, try to fit the parameters $A$ and $B$ such that $f(x) \approx A + Bx$.
    \item If the maximum deviation is below some threshold $\delta$, $\max_{x\in \Delta x_j}(\lvert f(x) - (A + Bx)\rvert) < \delta$, keep the interval $\Delta x_j$ and return. Otherwise, split $\Delta x_j$ in the middle, and recursively apply the same procedure for the left and right half separately.
\end{itemize}
Starting the above iteration with $\Delta x_j$ being the whole interval of $x$ results in the desired segmentation. Because of the specific construction of the intervals, in each recursive call, all $x$ belonging to the same $\Delta x_j$ have a common prefix of $q_j$ (highest significance) qubits (corresponding to the recursion depth) in their bitstring representation, whose pattern we call $\vec{q}_j$. For example, $\vec{q}_j = (1, 0, 0, 1, 0)$ (most significant bit on the left) means that all states in the interval $\Delta x_j$ start with the bit pattern $10010$, followed by their varying lower significance bits. Therefore, the \emph{flag} gates in \cref{fig:mainCirc} are simple multi-(anti-)controlled Toffoli gates.

For the slightly different problem of creating the mapping $\ket{x}\ket{0} \mapsto \ket{x}(f(x) \ket{0} + \sqrt{1 - f(x)^2} \ket{1})$, the parametrised fit function changes from $A + Bx$ to $\cos(A + Bx)$; everything else can stay the same.

As a side note, one can generally quantify whether a function is suitable for piecewise approximation using the Lipschitz constant \cite{sohrab2003basic} of the function. A large Lipschitz constant may imply a poor piecewise approximation. Moreover, the number of segments $S$, the size of the quantum register $n$, and the tolerance in the approximation are implicitly related. One would intuitively expect that e.g. to get a higher accuracy, one needs a larger register and a smaller tolerance; then the number of segments that come out of the bisection iteration would also be larger.

\section{Tiling Catalyst Factory Towers}

Here, we provide discussion on `tiling' sequential applications of catalyst tower circuits to reduce idle qubits and reduce depth by a factor of 2 in the limit of a large number of layers $n$.

\cref{fig:sequential_towers_normal} shows the sequential application of catalyst tower circuits, resulting in wasted space in the central region as the circuit must wait for the uncomputation on qubits $0,1$ before beginning another application. We provide an alternative, shown in \cref{fig:sequential_towers_tiled}, that presents the application of towers, now with the `orientations' reversed. The qubits at the top of the tower can now immediately begin execution of the second circuit upon being freed. In this case, each layer of the tower now requires an additional catalyst qubit, as the rotation angles differ in the up and down orientations.

\begin{figure*}
\centering
\begin{tikzpicture}
\begin{yquant}
nobit top;
qubit {$\ket{\psi_0}$} t00;
qubit {$\ket{\psi_1}$} t01;
nobit t10;
qubit {$\ket{\psi_2}$} t11;
nobit t20;
qubit {$\ket{\psi_3}$} t21; 
nobit bottom;

[name=ct0]
box {$\qquad\qquad\qquad\qquad CT\qquad\qquad\qquad\qquad$} (t00,t01);
hspace {10mm} t10;
align t10, t20, t21, bottom;
[name=ct1]
box {$\qquad\qquad\quad CT\qquad\qquad\quad$} (t10,t11);
align t20,t21, bottom;
hspace {6mm} t21;
[name=ct2]
box {$\qquad\quad CT \qquad\quad$} (t20,t21);
hspace {11mm} bottom;
[name=bottomrot]
box {$R_z(2^{k+2}\theta)$} bottom;

align -;

[name=ct0_next]
box {$\qquad\qquad\qquad\qquad CT\qquad\qquad\qquad\qquad$} (t00,t01);
hspace {10mm} t10;
align t10, t20, t21, bottom;
[name=ct1_next]
box {$\qquad\qquad\quad CT\qquad\qquad\quad$} (t10,t11);
align t20,t21, bottom;
hspace {6mm} t21;
[name=ct2_next]
box {$\qquad\quad CT \qquad\quad$} (t20,t21);
hspace {11mm} bottom;
[name=bottomrot_next]
box {$R_z(2^{k+2}\theta)$} bottom;

\draw ($(ct1.north west)+(0,-0.13)$) -| +(-0.3,0.3) ;
\draw ($(ct1.north east)+(0,-0.13)$) -| +(0.3,0.3) ;

\draw ($(ct2.north west)+(0,-0.13)$) -| +(-0.3,0.3) ;
\draw ($(ct2.north east)+(0,-0.13)$) -| +(0.3,0.3) ;

\draw ($(bottomrot.north west)+(0,-0.13)$) -| +(-0.3,0.3) ;
\draw ($(bottomrot.north east)+(0,-0.13)$) -| +(0.3,0.3) ;

\draw ($(ct1_next.north west)+(0,-0.13)$) -| +(-0.3,0.3) ;
\draw ($(ct1_next.north east)+(0,-0.13)$) -| +(0.3,0.3) ;

\draw ($(ct2_next.north west)+(0,-0.13)$) -| +(-0.3,0.3) ;
\draw ($(ct2_next.north east)+(0,-0.13)$) -| +(0.3,0.3) ;

\draw ($(bottomrot_next.north west)+(0,-0.13)$) -| +(-0.3,0.3) ;
\draw ($(bottomrot_next.north east)+(0,-0.13)$) -| +(0.3,0.3) ;

\draw [red,dotted, thick] ($(ct0.east)!0.5!(ct0_next.west)+(0,-3)$) -- +(0, 4) ;

\end{yquant}
\end{tikzpicture}
\caption{Two applications of a 3-layer tower, with wasted space in the center where the next tower is waiting for the previous tower to fully complete. The red dotted line indicates an undrawn layer of swapping out resource states and swapping in new states to have rotations applied. Catalyst states are not drawn for simplicity.}
\label{fig:sequential_towers_normal}
\end{figure*}
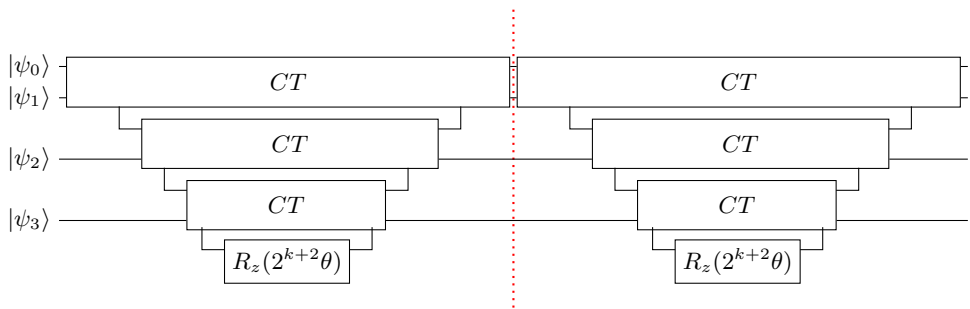

\begin{figure*}
\centering
\begin{tikzpicture}
\begin{yquant}
nobit top;
qubit {$\ket{\psi_0}$} t00;
qubit {$\ket{\psi_1}$} t01;
nobit t10;
qubit {$\ket{\psi_2}$} t11;
nobit t20;
qubit {$\ket{\psi_3}$} t21; 
nobit bottom;

[name=ct0]
box {$\qquad\qquad\qquad\qquad CT\qquad\qquad\qquad\qquad$} (t00,t01);
hspace {11mm} t10;
align t10, t20, t21, bottom;
[name=ct1]
box {$\qquad\qquad\quad CT\qquad\qquad\quad$} (t10,t11);
align t20,t21, bottom;
hspace {6mm} t21;
[name=ct2]
box {$\qquad\quad CT \qquad\quad$} (t20,t21);
hspace {11mm} bottom;
[name=bottomrot]
box {$R_z(2^{k+2}\theta)$} bottom;

hspace {4mm} t00,t01;
hspace {4mm} t10;
hspace {4mm} t21, t20;
hspace {2.7mm} t01;
hspace {8mm} t11;
discard t01;
discard t11;
init {} t20;
init {} t10;
hspace {4mm} t00;
hspace {4mm} t10;
hspace {4mm} t20, t21;

[name=ctinv0]
box {$\qquad\qquad\qquad\qquad CT\qquad\qquad\qquad\qquad$} (t21,t20);
hspace {1mm} t10, t11;
[name=ctinv1]
box {$\qquad\qquad\quad CT\qquad\qquad\quad$} (t10,t11);
hspace {4mm} t00, t01;
align t00, t01, top;
[name=ctinv2]
box {$\qquad\quad CT \qquad\quad$} (t00,t01);
hspace {5mm} top;
[name=toprot]
box {$R_z(2^{k+2}\theta)$} top;

\draw ($(ct1.north west)+(0,-0.13)$) -| +(-0.3,0.3) ;
\draw ($(ct1.north east)+(0,-0.13)$) -| +(0.3,0.3) ;

\draw ($(ct2.north west)+(0,-0.13)$) -| +(-0.3,0.3) ;
\draw ($(ct2.north east)+(0,-0.13)$) -| +(0.3,0.3) ;

\draw ($(bottomrot.north west)+(0,-0.13)$) -| +(-0.3,0.3) ;
\draw ($(bottomrot.north east)+(0,-0.13)$) -| +(0.3,0.3) ;

\draw [red,dotted, thick] ($(ct2.east)!0.5!(ctinv0.west)+(0,-0.42)$) -- +(0, 0.84) ;

\draw [red,dotted, thick] ($(ct2.east)!0.5!(ctinv0.west)+(0,0.42)$) -- ($(ct1.east)!0.5!(ctinv1.west)+(0,-0.42)$);

\draw [red,dotted, thick] ($(ct1.east)!0.5!(ctinv1.west)+(0,-0.42)$) -- +(0, 0.84) ;

\draw [red,dotted, thick] ($(ct1.east)!0.5!(ctinv1.west)+(0,0.42)$) -- ($(ct0.east)!0.5!(ctinv2.west)+(0,-0.42)$);

\draw [red,dotted, thick] ($(ct0.east)!0.5!(ctinv2.west)+(0,-0.42)$) -- +(0, 0.84) ;

\draw ($(ctinv1.south west)+(0,0.13)$) -| +(-0.3,-0.3) ;
\draw ($(ctinv1.south east)+(0,0.13)$) -| +(0.3,-0.3) ;

\draw ($(ctinv2.south west)+(0,0.13)$) -| +(-0.3,-0.3) ;
\draw ($(ctinv2.south east)+(0,0.13)$) -| +(0.3,-0.3) ;

\draw ($(toprot.south west)+(0,0.13)$) -| +(-0.3,-0.3) ;
\draw ($(toprot.south east)+(0,0.13)$) -| +(0.3,-0.3) ;

\end{yquant}
\end{tikzpicture}
\caption{Two applications of a 3-layer tower tiled to reduce qubit idling. The red dotted line indicates an undrawn layer of swapping out resource states and swapping in new states to have rotations applied. In this case, two catalyst states are required for each layer of the tower, one for each of the `up' and `down' orientations.}
\label{fig:sequential_towers_tiled}
\end{figure*}
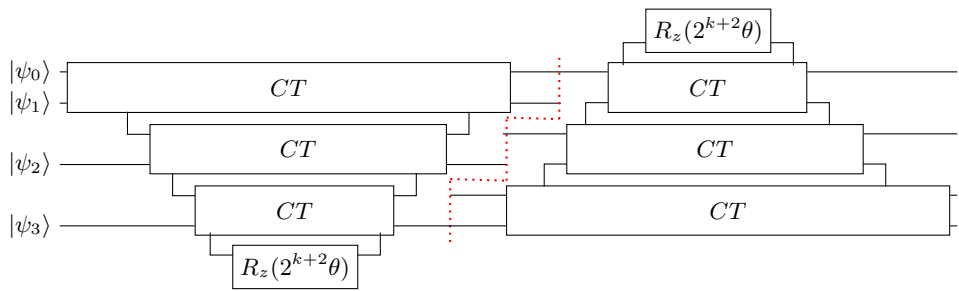

\section{Expected Depth of Parallel Repeat-Until-Success Rotations} \label{appx:depth}

As described in \cref{sec:options}, rotations can be implemented using resource states in a repeat-until-success procedure \cite{BocharovRoettelerSvore2014}, where gate teleportation succeeds with a probability of $0.5$, and if failure occurs, another attempt is made with twice the angle: to fix the incorrect angle from the previous teleportation(s) and apply the desired rotation.
For a single rotation, the expected depth of this procedure is $2$, but when performing $m$ of these rotations in parallel as in \cref{fig:mainCirc}, we must wait for all the rotations to have finished before applying subsequent multi-qubit gates. We therefore must account for the possibility that some of the rotations have a large number of failures, and the expected depth increases as we increase $m$.

Given the random variables $D_1$ and $D_m$ which are the depth of the repeat-until-success procedure for $1$ rotation and $m$ rotations respectively, we can write the probability that $D_m \le d$ as:

\begin{equation}
P(D_m \le d) {=} P(D_1 \le d)^m \\
{=}(1-P(D_1 > d))^m\\
{=} (1-2^{-d})^m
\end{equation}
as we simply require that all $m$ rotations have terminated by depth $d$.

We can then find the probability that the depth for $m$ rotations is exactly $d$ by using this cumulative probability to find the probability density:

\begin{align}
    P(D_m=d) =P(D_m \le d) - P(D_m \le d-1)
\end{align}

Resulting in the expected depth:

\begin{equation}
    \mathbb{E}[D_m] = \sum_{d=1}^\infty d((1-2^{-d})^m- (1-2^{-d+1})^m)
\end{equation}

\section{Comments on Linear interpolation via QROM} \label{appx:qrom_comments}

Here we provide further comments on the application of a phase of a linear interpolation of a function $\tilde f(x) = a_i x +b_i$, where the slope and intercept $a_i,b_i$ vary with $x$ via QROM and arithmetic functions \cite{sandersCompilationFaultTolerantQuantum2020}. The calculation of $\ket{\tilde f(x)}$ can then be performed with a single multiplication and addition, with the phase applied via addition into a phase gradient state.

We begin by making choices for the addition and multiplication methods used to calculate $\tilde f(x)$, choosing Brent-Kung carry look-ahead addition using logical-AND gates where possible as described in \cite{wangOptimalToffoliDepthQuantum2024}, and the Karatsuba multiplication method from Jang et al. \cite{jangQuantumBinaryField2023} which has the lowest known depth of quantum multiplication.
Taking the addition, and converting from the Toffoli depth provided in \cite{wangOptimalToffoliDepthQuantum2024}, we need $3\lceil \log_2 n  \rceil-1$ measurement depth and $\sim 22n$ T count. The multiplication comes to a measurement depth of $4$ (ignoring the log-depth Clifford circuits that are a component of this circuit) and the number of $T$-gates required for level-$p$ Karatsuba multiplication is $7 \cdot 3^p (n/2^p )^2$, which for $n$ a power of $2$ and fully parallelised to $p=\log_2(n)$ is $7 \cdot 3^{\log_2(n)}$ (this expression is only exact when $n$ is a power of $2$, otherwise $7 \cdot3^{\lceil \log_2(n)\rceil}$ can be used as an upper bound) \cite{jangQuantumBinaryField2023}. We note that this implementation of Karatsuba multiplication also requires a number of ancilla qubits scaling as $3\cdot 3^{\log_2n}$ (this method with its $O(n^{\log_2(3)})$ qubit count scaling is the only one presented in \cref{tab:expressions} with a super-linear scaling in $n$). Overall, we find a measurement depth of $9\lceil \log_2 n \rceil +1$ for the computation and uncomputation of $\tilde f(x)$, along with the central phase-gradient addition. We can therefore conclude that even when using the most depth-optimised multiplication circuits, the depth of this method is outperformed by the injection of resource states prepared by independent catalyst towers in the regime where a simple piecewise approximation of the function is applicable.

The overall $T$ costs for the phase oracle using this circuit are given by two applications of multiplication and addition, addition with the phase gradient state, and the original creation of the phase gradient state which requires $n$ single qubit rotations.

We also note that the parallelisation of this method will result in resource state savings when compared to the catalyst tower method described in \cref{sec:options}, as the data from QROM output can be placed onto a single register, even when computed in parallel, meaning that the arithmetic and application of phase can be performed only once, resulting in resource state savings, and only requiring one phase gradient state. See \cref{fig:qrom_parallel} for an example circuit, with the SelectSwap method of \cite{krishnakumarImplementationQuantumLookup2022} being an alternative for the parallel loading of the $a_i, b_i$, requiring fewer ancilla qubits but leaving many of them dirty for the central section of the circuit.

A further optimisation is possible to improve the scaling of the central multiplication. The register $x$ can take $2^n$ discrete values, but when the piecewise decomposition of the function has a largest segment of size $\Delta x_L$ containing $2^q$ values it is instead possible to perform the multiplication only against the \emph{offset} from the start of this segment, with cost determined by the largest segment as $O(7\cdot3^{\log_2(q)})$, at the expense of performing extra additions and loading the start locations via QROM (although this can in some cases be avoided if the segments follow convenient patterns, such as following powers of two as described in \cite{sandersCompilationFaultTolerantQuantum2020}). However, functions that are desirable to implement in a piecewise fashion often have largest segments that take up a large fraction of the total domain as in \cref{fig:linear_derivatives,fig:coulomb_potential_approx}, so the possible improvement should be weighed against the slight additional cost from the extra addition in $a_i'(x-s_i)+b_i'$, where $s_i$ is the value where the $i$th segment begins.


%

\end{document}